%% file: main.tex
\documentclass{article}
\PassOptionsToPackage{numbers,sort&compress}{natbib}
\usepackage[final,sglblindworkshop]{neurips_2026}
\workshoptitle{Foundation Models for Temporal Systems}
\usepackage[utf8]{inputenc}
\usepackage[T1]{fontenc}
\usepackage{booktabs,amsmath,microtype,graphicx,url,longtable}
\usepackage[hidelinks]{hyperref}

\input{extension_numbers}

\input{ladder_numbers}
\title{Where Should the Code Go?\\Locating Foundation-Model Errors in Bitemporal Inventory Simulation}
\author{Durga Sandeep Saluru\\
Distyl AI\\
\texttt{durga@distyl.ai}
\And
Lanchen Song\\
Distyl AI\\
\texttt{lanchen.song@distyl.ai}
\And
Aditya Kumaran\\
Distyl AI\\
\texttt{aditya.kumaran@distyl.ai}}
\begin{document}
\maketitle
\begin{abstract}
Operational reports carry two clocks: when an event happened and when its report arrived;
later reports can revise earlier ones. A foundation-model component that maintains the
state supported by the reports visible at a cutoff, then rolls a supplied scenario forward,
can be helped by exact code before it, after it or inside the call, or asked to reason harder. We ask
which remedy removes which error, for which models, on an exactly solvable lost-sales
inventory environment: \TotalRequests\ independently rescored requests over twelve
configurations of closed and open-weight models, with a forced-answer-call interface as a
control and protocols frozen before the calls. Weak models fail in separable ways. Canonical input removes
report-selection error, exact continuation removes rollout error, neither repairs the other,
and which error dominates differs by model; strong models are at ceiling on raw reports.
Asking for more reasoning is no substitute and can depend on the interface: under a forced
answer call high effort collapses GPT-5.4 mini and DeepSeek V4 Pro, which gain or hold under
automatic choice; on its provider the forced call disables reasoning for DeepSeek V4 Flash;
and at high effort gpt-oss exhausts its output budget. The contribution is a controlled
component diagnostic, released with every request, and evidence that for some model and
provider pairs the tool-calling interface confounds reasoning evaluation.
\end{abstract}

\section{Motivation and related work}
The state supported by the reports visible at a knowledge cutoff $K$ is
therefore not the physical state, and reconstructing it requires an arrival-time cutoff,
revision resolution, and replay in event-time order. Lost-sales dynamics make order
consequential: unmet demand is lost, so swapping a receipt and a demand changes the
trajectory at equal totals. We study a component that maintains this temporal state from an
event stream and then applies an explicitly supplied future schedule, which separates state
error from forecasting uncertainty and makes the reporting cutoff a leakage control rather
than a nuisance.

Ordinary code solves this task exactly, yet models hold operational state in agentic
pipelines: inventory agents reason over order and arrival histories in
context~\citep{baek2026inventory}, and their stated beliefs and actions can
diverge~\citep{deb2026stocktake}. We ask where the designer of such a pipeline should put
exact code, and which placement removes which error for which models, when every
intermediate quantity has an executable ground truth; the seeded environment and every
request and response are released.\footnote{\url{https://github.com/durga-sandeep/where-should-the-code-go}}

\paragraph{Related work.}
Temporal QA benchmarks test when facts hold (valid
time)~\citep{chu2024timebench,luo2025etrqa,wei2025time,weng2026temporalbench,tdbench2025,evolveqa2025}; BiTempQA annotates
event and record time for QA~\citep{bitempqa}. None asks a model to reconstruct a numeric
state \emph{as of} a record time under revisions, and the inventory benchmarks that do use
lost-sales dynamics evaluate ordering policy, not ledger replay~\citep{baek2026inventory,deb2026stocktake}.
Program-aided prompting puts exact execution after the model~\citep{gao2023pal,chen2023pot}
and code-interpreter agents put it inside the call, trading arithmetic error for other
errors~\citep{toolmyopia2025}; the three-way comparison with per-stage error attribution is,
to our knowledge, new. On interfaces, format
constraints are known to tax reasoning, most for weaker
models~\citep{tam2024speakfreely,banerjee2025crane,fan2026capacity}, and more test-time
reasoning can lower accuracy~\citep{gema2025inverse}; we find that for some model and
provider pairs (clearly GPT-5.4 mini and DeepSeek V4 Pro, slightly GPT-5.4) the format tax
grows with requested effort under a forced answer call, while for four others it does not. State tracking degrades with depth across
scales~\citep{wmfam2026,sinha2025illusion}; our model ladder is only the substrate here.

\section{Environment and study design}
\paragraph{Semantics.}
A report has a logical event identifier, revision, event time, received time, kind and
quantity. For target time $T$ and cutoff $K$: keep reports received by $K$, retain the
highest visible revision per event, keep events with event time at most $T$, and replay
in (event time, identifier) order from the initial stock. A receipt of $q$ adds $q$; a
demand of $q$ sets $x'=\max(0,x-q)$ and loses $\max(0,q-x)$. A supplied schedule through
horizon $H$ is then applied with the same rule. The full output is five integers: stock at
$T$, historical loss, stock at $H$, future loss, and the first future time with positive
loss ($-1$ if none; formal rule and worked example in Appendix~\ref{app:semantics}).

\paragraph{Where the code sits.}
Three full-output arms move executable work before the model. \textbf{Raw} supplies every
report version in received-time order, including reports after $K$ that must be ignored;
\textbf{canonical} supplies the correctly selected, version-resolved events in event-time
order; \textbf{snapshot} additionally supplies the exactly replayed historical state, so
only the rollout remains. Two \textbf{state-only} arms move executable work after the
model: with raw or canonical input and no schedule, the model returns only stock at $T$ and
historical loss, and code applies the schedule to the emitted stock, giving a five-field
\emph{hybrid} answer. Two \textbf{code-inside} arms give the OpenAI configurations a hosted code tool, with raw or
canonical input and the full output; the model chooses whether to run code. A post-hoc \textbf{continuation} applies the same code to the stock emitted in the full-output
arms; by conservation of units it is exact exactly when that stock is, so hybrid future
accuracy equals reconstruction accuracy (an identity, verified on every answer).

\paragraph{Configurations, cohorts and statistics.}
Twelve configurations: GPT-5.4 mini and GPT-5.4 (OpenAI), Claude Haiku~4.5 and Sonnet~4.6
each without and with extended thinking (Anthropic), and six open-weight models through
OpenRouter, each pinned to one upstream provider: gpt-oss-20b and gpt-oss-120b, Qwen3.8-27B,
DeepSeek V4 Flash and V4 Pro, and GLM-5.3 Flash (3.6B to 49B active parameters). The main
table uses automatic tool choice at the lowest requested effort for every model; the OpenAI
and open-weight models are also run under a forced answer call (the answer function is the
only permitted action) at requested effort low and high, giving a $2\times2$ effort by
interface cross per model. Requests use a strict integer tool schema, 16,384 output tokens and each provider's default
sampling (we set no temperature or top-$p$), with no retries of any model output; missing, malformed, refused or truncated outputs fail.
The synthetic cohort has 96 seeded heldout histories in a 24-cell factorial over history length, reporting
delay, corrections and schedule type (Appendix~\ref{app:semantics}). The retail-profile cohort maps 48 public
FreshRetailNet~\citep{wang2025freshretailnet} store windows to demand quantities; stock,
receipts, delays and revisions are simulated. Each configuration/arm/history cell has two
requests. Endpoints are all-five exactness and \emph{future-joint} exactness (all three
scenario outputs). Paired differences average repetitions within history and use 10,000
history-bootstrap percentile intervals, reported as descriptive 95\% intervals without
multiplicity adjustment. They do not capture variation between collection periods: re-running
one identical cell several days later moved it by 8.9 points, so we treat smaller
differences as unresolved. Every study was fixed in a written protocol before its calls;
independent re-parsing and re-scoring of every response found zero mismatches
(Appendices~\ref{app:extensions} and~\ref{app:ladder}; controls in Appendix~\ref{app:controls}).

\section{Results}
\input{ladder_table}
\paragraph{Code before the model removes selection error and most replay error, not rollout error.}
\input{ladder_placement_findings}

\paragraph{Code after the model removes rollout error but cannot repair state.}
This holds by construction (Section~2); what is empirical is how much each model gains.
\input{ladder_after_findings}

\paragraph{Code inside the call.}
\input{extension_tool_findings}

\paragraph{Reasoning harder fails in three interface-dependent ways.}
\input{interaction_table}
\input{ladder_interface_findings}

\paragraph{State and decision errors differ (post hoc).} For mini, 166 of 208 wrong starting
stocks still give the right stockout decision (Appendix~\ref{app:diagnostics}).

\section{Limitations and implications}
The task is solved exactly by code; the results say where to place code around a model that
already holds operational state, not whether a model should replace code. Synthetic histories, artificial corrections, two repetitions, one retail source
and 96 histories limit coverage; differences under nine points, including GLM's canonical gain and
Haiku's hybrid gain, are not resolved. Snapshot input is not a pure rollout probe for every
model: GLM, and DeepSeek V4 Pro on retail profiles, score lower with exact state than with
canonical input, so the state's presentation may itself change behaviour.
Each open-weight model is served by one upstream provider whose quantization and effort
mapping we do not control, so its interface effects are claims about the model and provider
pair, not the model; missing answers, refusals and truncations count as failures,
so some cells (gpt-oss-20b, high-effort gpt-oss-120b) measure completion as much as
accuracy. GPT-5.4 under automatic choice scored lower on the snapshot and hybrid arms than
under the forced call collected a day earlier while its raw and canonical cells did not move;
\GptReplicationSentence
What the diagnostic supports is narrower than its title: for weak components, reconstruction
and rollout fail separately and code can be placed to remove each; for strong components the
placement is moot; and an evaluation that raises reasoning effort through a forced tool call,
or compares families under different tool-calling interfaces, can measure the interface as
much as the model, as it clearly did here for three of the eight pairs crossed.

\paragraph{Pipelines with a time-series foundation model (TSFM).}
No TSFM is evaluated; forecasting was removed to measure state error exactly. Upstream, when an LLM reconstructs the history a TSFM consumes from late and revised reports,
selection and replay errors become forecast inputs no forecaster can repair; normalizing
the reports in code first removes them. Downstream, when a TSFM forecast supplies the
schedule, the rollout our snapshot and hybrid arms isolate belongs in code, so rollout error
is not mistaken for forecast error. The cutoff $K$ is also the leakage control a forecasting
backtest needs: scoring on revisions not visible at $K$ overstates accuracy.

\paragraph{AI assistance.} AI assistants were used extensively in this work; the authors are responsible for all claims.

\clearpage
\bibliographystyle{plainnat}
\bibliography{references}
\clearpage
\input{appendix}

\end{document}

%% file: extension_numbers.tex
\newcommand{\TotalRequests}{36,288}
\newcommand{\EffortRequests}{3,648}
\newcommand{\StateRequests}{2,880}
\newcommand{\ToolRequests}{1,152}
\newcommand{\ReplicationRequests}{576}

%% file: ladder_numbers.tex
\newcommand{\GptReplicationSentence}{a same-period forced-call replication of those two arms (Appendix~\ref{app:ladder}) moved them by -1.0 and +0.5 points relative to the original cells, so the automatic-choice difference against the contemporaneous baseline is -6.8 and -15.1 points.}

\renewcommand{\TotalRequests}{36,672}

%% file: ladder_table.tex
\begin{table}[t]
\centering\footnotesize
\setlength{\tabcolsep}{3.5pt}
\begin{tabular}{lrrrrrrr}
\toprule
 & \multicolumn{4}{c}{Synthetic} & \multicolumn{3}{c}{Retail profiles} \\
\cmidrule(lr){2-5}\cmidrule(lr){6-8}
Model (automatic tool choice) & raw & canonical & hybrid & snapshot & raw & canonical & snapshot \\
\midrule
GPT-5.4 mini & 56.8 & 86.5 & 94.8 & 94.8 & 14.6 & 82.3 & 89.6 \\
Haiku 4.5 & 52.6 & 85.9 & 92.7 & 92.7 & 6.2 & 89.6 & 92.7 \\
Haiku 4.5 + thinking & 88.5 & 96.9 & 100.0 & 99.0 & 40.6 & 91.7 & 96.9 \\
gpt-oss-20b$^\dagger$ & 24.0 & 51.0 & 71.4 & 82.8 & 0.0 & 17.7 & 77.1 \\
gpt-oss-120b$^\dagger$ & 49.5 & 73.4 & 76.0 & 94.8 & 5.2 & 72.9 & 86.5 \\
Qwen3.8-27B$^\dagger$ & 95.3 & 99.5 & 100.0 & 100.0 & 96.9 & 99.0 & 100.0 \\
DeepSeek V4 Flash$^\dagger$ & 98.4 & 99.0 & 99.0 & 99.0 & 94.8 & 96.9 & 99.0 \\
GLM-5.3 Flash$^\dagger$ & 55.7 & 64.1 & 87.0 & 58.9 & 26.0 & 58.3 & 38.5 \\
DeepSeek V4 Pro$^\dagger$ & 99.0 & 99.0 & 99.0 & 99.0 & 95.8 & 100.0 & 95.8 \\
GPT-5.4 & 98.4 & 98.4 & 83.9 & 91.1 & 97.9 & 100.0 & 89.6 \\
Sonnet 4.6 & 99.0 & 99.5 & 100.0 & 100.0 & 95.8 & 99.0 & 100.0 \\
Sonnet 4.6 + thinking & 97.9 & 100.0 & 100.0 & 100.0 & 97.9 & 100.0 & 100.0 \\
\bottomrule
\end{tabular}
\caption{Future-joint accuracy (\%, all three scenario outputs exact) by input and code placement, every configuration under automatic tool choice at its lowest requested effort (OpenAI and open-weight models: requested effort low; Anthropic models: no or 4,096-token thinking); 192 synthetic and 96 profile calls per cell; invalid and missing answers fail. \emph{raw}: model selects, replays and rolls out; \emph{canonical}: code selects and orders eligible events; \emph{hybrid}: model returns the two historical fields from canonical input and code applies the schedule (equals reconstruction accuracy by conservation); \emph{snapshot}: exact state supplied, model rolls out. $^\dagger$Open-weight models through OpenRouter, each pinned to one upstream provider (Appendix~\ref{app:ladder}). Forced-call cells and all-five accuracy: Appendix~\ref{app:tables}.}
\label{tab:ladder}
\end{table}

%% file: ladder_placement_findings.tex
Table~\ref{tab:ladder} reports every configuration under one interface; 6 of 12 start below 90\% on raw reports and the rest are near ceiling there, so placement is a question about weak components. For the weak ones, canonical input raises future-joint accuracy by mini +29.7 [+20.8, +38.5]; Haiku +33.3 [+24.0, +43.2]; Haiku+thinking +8.3 [+3.1, +14.1]; gpt-oss-20b +27.1 [+19.8, +34.4]; gpt-oss-120b +24.0 [+16.1, +31.8]. Exact state does not make them exact: with only the 12-event rollout left, every weak model stays below 100\% on snapshot input (Table~\ref{tab:ladder}). The decomposition also locates different errors in different models. For GLM canonical input adds +8.3 [+0.0, +16.7] and exact state adds nothing over it (-5.2 [-14.6, +4.7]), yet code after the model on the same canonical answers adds +22.9 [+15.1, +30.7]: its error is almost entirely rollout. For gpt-oss-20b both stages fail (+27.1 [+19.8, +34.4] from canonical input, +31.8 [+21.4, +42.2] more from exact state).

%% file: ladder_after_findings.tex
Run prospectively, the state-only hybrid's gain over the model's own rollout on canonical input is mini +8.3 [+3.6, +13.5]; Haiku +6.8 [+0.0, +13.5]; gpt-oss-20b +20.3 [+12.0, +28.6]; gpt-oss-120b +2.6 [-4.2, +9.4]; GLM +22.9 [+15.1, +30.7]. On raw input the hybrid never exceeds 78\% for a weak model: code after the model removes rollout error wherever the reconstruction was right and cannot repair a stock that is wrong because the wrong reports were selected.

%% file: extension_tool_findings.tex
With a hosted code-execution tool available (OpenAI configurations only), mini's raw-input future-joint accuracy rises by +54.2 [+45.3, +62.5] points on synthetic histories and +77.1 [+68.8, +84.4] on profiles, and GPT-5.4 reaches 100\% on raw input in both cohorts; mini ran code in 469 of 576 calls and was exact in 453 of them, versus 93 of the 107 calls it answered without code. Against the interface-matched automatic-choice baseline the tool's own contribution on synthetic histories is +34.9 [+26.0, +43.8] points on raw and +11.5 [+6.2, +17.7] on canonical input. Letting the model delegate selection and rollout to code it writes is the most effective single placement at fixed input, though not error-free (Appendix~\ref{app:extensions}).

%% file: interaction_table.tex
\begin{table}[t]
\centering\footnotesize
\setlength{\tabcolsep}{3.5pt}
\begin{tabular}{lrrrrrrr}
\toprule
 & \multicolumn{3}{c}{Forced answer call} & \multicolumn{3}{c}{Automatic tool choice} & \\
\cmidrule(lr){2-4}\cmidrule(lr){5-7}
Model & low & high & high$-$low & low & high & high$-$low & Interaction \\
\midrule
GPT-5.4 mini & 76.0 & 34.9 & -41.1 & 86.5 & 93.2 & +6.8 & +47.9 [+35.9, +59.9] \\
GPT-5.4 & 98.4 & 94.8 & -3.6 & 98.4 & 100.0 & +1.6 & +5.2 [+1.0, +9.4] \\
gpt-oss-20b & 52.1 & 66.1 & +14.1 & 51.0 & 59.4 & +8.3 & -5.7 [-15.6, +4.2] \\
gpt-oss-120b & 69.3 & 66.1 & -3.1 & 73.4 & 68.8 & -4.7 & -1.6 [-10.4, +6.8] \\
Qwen3.8-27B & 91.7 & 91.7 & +0.0 & 99.5 & 92.7 & -6.8 & -6.8 [-14.1, +0.5] \\
DeepSeek V4 Flash & 2.1 & 1.0 & -1.0 & 99.0 & 97.4 & -1.6 & -0.5 [-4.7, +3.1] \\
GLM-5.3 Flash & 62.5 & 82.3 & +19.8 & 64.1 & 84.4 & +20.3 & +0.5 [-9.9, +11.5] \\
DeepSeek V4 Pro & 98.4 & 82.3 & -16.1 & 99.0 & 97.4 & -1.6 & +14.6 [+8.3, +20.8] \\
\bottomrule
\end{tabular}
\caption{Requested reasoning effort $\times$ tool-choice interface, synthetic cohort, canonical input, future-joint accuracy (\%; 192 calls per cell). high$-$low: paired effort effect within an interface. Interaction: automatic-choice effort effect minus forced-call effort effect, paired by history, with a descriptive 95\% history-bootstrap interval (10,000 draws); a positive interaction means the forced call turns extra effort into a liability. Requested effort is a family-specific provider setting, not a compute match (observed reasoning tokens: Appendix~\ref{app:ladder}).}
\label{tab:interaction_canonical}
\end{table}

%% file: ladder_interface_findings.tex
Table~\ref{tab:interaction_canonical} crosses requested effort with the tool-calling interface. For DeepSeek V4 Flash on its provider the forced call effectively disabled reasoning: forced canonical calls averaged 32 reasoning tokens against 1695 under automatic choice, and accuracy is 2.1 against 99.0\%; effort changes nothing under either interface. For GPT-5.4 mini and for DeepSeek V4 Pro on its provider the forced call interacts with effort: mini's high-minus-low effect on canonical input is -41.1 [-51.0, -31.2] under the forced call and +6.8 [+0.5, +13.5] under automatic choice, and DeepSeek V4 Pro falls from 98.4 to 82.3\% under the forced call at high effort (32 of 192 calls with no answer) while staying at 97.4\% under automatic choice. GPT-5.4 shows a small interaction of the same sign (+5.2 [+1.0, +9.4]). For gpt-oss-20b, gpt-oss-120b, Qwen3.8-27B and GLM-5.3 Flash the interaction interval covers zero: effort acts the same way under both interfaces. A third failure of \emph{reason harder} is budget exhaustion: at high effort gpt-oss-120b spent the whole 16,384-token output limit on reasoning in 65 of 192 forced canonical calls and gpt-oss-20b in 97 of 192 automatic raw calls, each a missing answer. gpt-oss-20b also refused the fictional task (88 of 192 low-effort raw calls had no answer, mostly refusals), so its cells measure willingness as much as ability. Reasoning tokens and missing answers for every cell: Appendix~\ref{app:ladder}.

%% file: appendix.tex
\appendix
\section{Executable semantics, provider settings and sampling}
\label{app:semantics}
The task reconstructs the state supported by a versioned report history, rather
than recovering unobserved physical inventory. For each report $r$, let $e(r)$ be
its logical event identifier, $v(r)$ its revision, $t(r)$ its event time and
$a(r)$ its received time. The selected version is
\[
 r_e^*(K)=\arg\max_{r:\,e(r)=e,\,a(r)\leq K} v(r).
\]
Identifiers with no visible report are absent. Only selected versions with
$t(r_e^*)\leq T$ are replayed. Thus a newer visible revision outside the target
horizon does not cause an older revision to be substituted. This edge case is
covered by unit tests, although generated experimental historical event times
all precede $T$. The generator uses distinct revisions, so revision ties do not
arise. Replay sorts by event time and then event identifier.

\paragraph{Illustrative two-clock example.}
Start with two units of stock. A receipt of five units occurs at event time 1,
but its report arrives at time 4. A demand of six units occurs and is reported
at time 2. Reconstruct through $T=3$. With $K=3$, only the demand is visible:
the reconstructed stock is zero and historical loss is four. With $K=4$, replay
the receipt before the demand: reconstructed stock is one and historical loss
is zero. A supplied future demand of three units at time 5 then loses three
units in the first view and two in the second. Both views finish at zero stock
and have first future stockout time 5. The later knowledge view revises an
earlier reconstruction; it is not a forecast that had access to later reports.

\paragraph{Resource and output contract.}
OpenAI requests use dated model identifiers
\texttt{gpt-5.4-mini-2026-03-17} and \texttt{gpt-5.4-2026-03-05}, requested
reasoning effort low (medium and high for the effort sweep), and forced
\texttt{submit\_answer} tool choice. Both Anthropic configurations use
\texttt{claude-sonnet-4-6}, automatic tool choice, and disabled parallel tool use;
one additionally requests manual extended thinking with a 4096-token budget. The
budget is a requested thinking target, not a strict observed token ceiling. Full
arms use a strict five-integer schema; state-only arms use a strict two-integer
schema and instructions that replace the future-scenario paragraph with an explicit
statement that no future is to be simulated. All requests allow 16384 total output
tokens. Automatic tool choice may yield reasoning text without an answer; absent or
malformed answers remain failures. Requests have a 420-second transport timeout and
are not retried or repaired. Exact payloads, returned token usage and provider
responses are included in the supplement. These interfaces and compute allowances
are disclosed rather than treated as a controlled cross-provider comparison.

\paragraph{Sampling and uncertainty.}
The primary cohort contains 2304 requests, the later-knowledge control 576,
the display-order control 192, and the profile cohort 1152, for 4224 in total.
The effort sweeps add \EffortRequests\ requests, the state-only arms \StateRequests, the
code-inside arms \ToolRequests, and the replication \ReplicationRequests, all on the same histories. The controls reuse the first 24 primary
histories. Each of the 96 primary histories and 48 profile windows has two
repetitions per configuration and arm. Repetition averages are paired within the
same history/window before bootstrap resampling. Intervals characterize variation
within this constructed sample; they are not population guarantees or corrections
for multiple comparisons. Source-store separation does not establish independence
of retail behavior across stores. Extension studies add configurations and arms on
the archived histories, not independent histories.
\input{coverage_text}
\paragraph{Overlap and contamination scope.}
A post-hoc exact-content audit finds no duplicates within, or overlap between,
24 regenerated v2 development histories and the 96 heldout histories, or between
16 development and 48 heldout profile tasks. Profile source stores are disjoint;
exact ordered quantized sales sequences also have no overlap. Checks use frozen
generators and are archived with their scope. Intentional control/repetition
reuse is excluded. These checks do not test near-duplicates or provider training
data; the public sales profiles may have appeared in training. No cases were
removed or rescored by this audit.
\paragraph{Retail-profile time axis.}
Source hours determine ordered two-hour sales bins. The adapter assigns synthetic
event times to alternating historical receipts/demands and interleaved future
receipts/demands; these event-time units are not literal calendar hours. The
profile check preserves sequential quantity patterns rather than validating
real operational reporting timestamps.

\section{All-five accuracy tables}
\label{app:tables}
\begin{figure}[htbp]
  \centering
  \includegraphics[width=\linewidth]{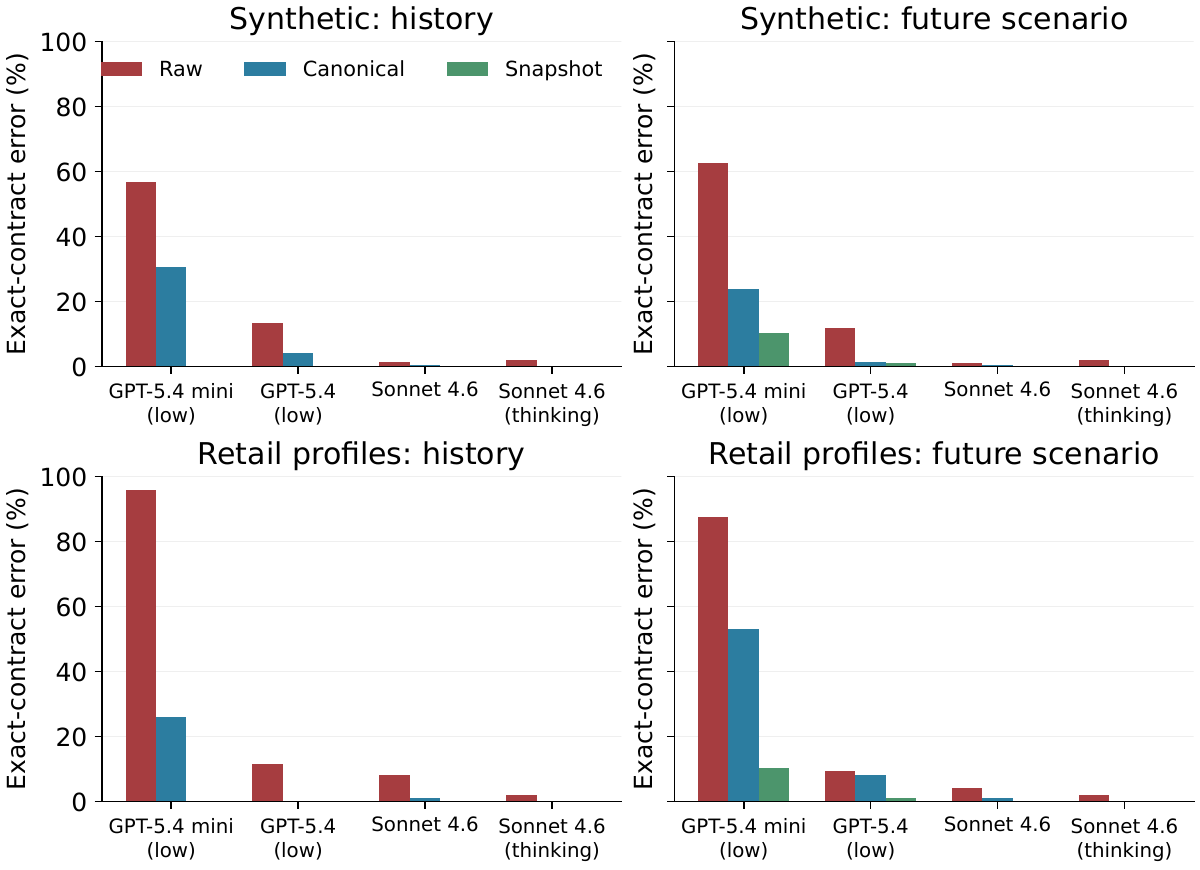}
  \caption{Historical and future-stage errors under the three representations in
  the original grid. Snapshot historical answers are supplied by code; their accuracy
  measures faithful copying. Future-stage errors remain the substantive endpoint for
  that representation.}
  \label{fig:errors}
\end{figure}

\begin{figure}[htbp]
  \centering
  \includegraphics[width=\linewidth]{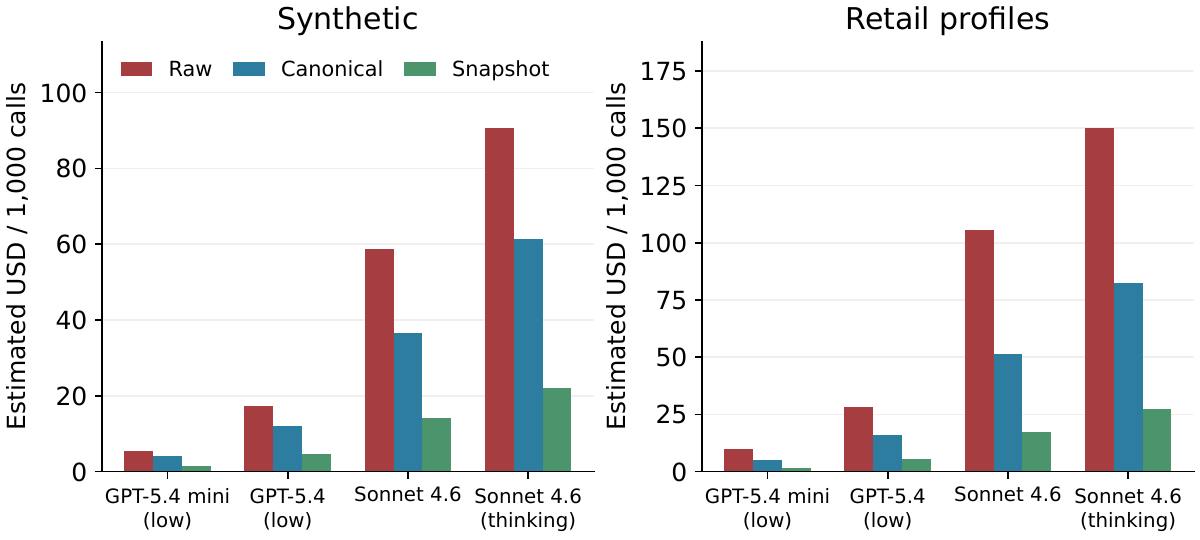}
  \caption{Observed API token-cost estimates under each representation in the
  original grid. These are configuration-dependent resource measurements, not
  compute-matched comparisons or provider billing invoices.}
  \label{fig:cost}
\end{figure}

Tables~\ref{tab:main} and~\ref{tab:external} report both the all-five and the
future-joint endpoints for the original grid; Table~\ref{tab:factors} stratifies mini's
synthetic results by design factor; Table~\ref{tab:allcells} lists every cell including
the extension studies. Figures~\ref{fig:errors} and~\ref{fig:cost} plot the original
grid's historical and future error rates and observed token cost by representation.
\input{main_table}
\input{external_table}
\input{factor_table}
\input{extension_all_cells_table}

\section{Paired temporal and display controls}
\label{app:controls}
On the first 24 heldout bases (one factorial cycle), a paired view changes only
$K$ to 130, revealing later reports. This is retrospective revised-state simulation,
not a forecast made at time 100. Another control shuffles raw report display order
while preserving all timestamps and the correct answer. We report both-view
correctness as well as sensitivity and invariance, since consistent wrong answers
are not evidence of reliability (Table~\ref{tab:controls}).
\input{controls_table}
Separate requests also vary stochastically; cross-view disagreement alone does
not isolate an order effect. Table~\ref{tab:repeat} uses the existing two repetitions on the same 24 control
bases. Within-view pairs compare repetitions 0 and 1; cross-view pairs compare
corresponding repetitions across the original and shuffled views. All are
separate API requests, with no shared random seed. Thus cross-view disagreement
includes ordinary request variation. Mini's cross-view agreement of 27.1\%
lies between its within-view rates of 25.0\% and 33.3\%; these observations do
not isolate a causal display-order effect. This diagnostic was added after
collection of the synthetic cohorts, without additional model calls.
\begin{table}[htbp]
\centering\small
\begin{tabular}{lrrr}
\toprule
Configuration & Original repeats & Shuffled repeats & Cross-view \\
\midrule
GPT-5.4 mini (low) & 25.0 / 20.8 & 33.3 / 33.3 & 27.1 / 25.0 \\
GPT-5.4 (low) & 70.8 / 70.8 & 70.8 / 70.8 & 68.8 / 68.8 \\
Sonnet 4.6 & 95.8 / 95.8 & 83.3 / 83.3 & 89.6 / 89.6 \\
Sonnet + thinking & 91.7 / 91.7 & 95.8 / 95.8 & 93.8 / 93.8 \\
\bottomrule
\end{tabular}
\caption{Post-hoc repeat diagnostics (\%): identical valid answers / both correct.
Within-view columns have 24 repetition pairs; cross-view has 48 matched-repetition
pairs over those same 24 bases. These are descriptive rates, not independent
samples of histories.}
\label{tab:repeat}
\end{table}

\section{Consistency and decision diagnostics}
\label{app:diagnostics}
\begin{figure}[htbp]
  \centering
  \includegraphics[width=\linewidth]{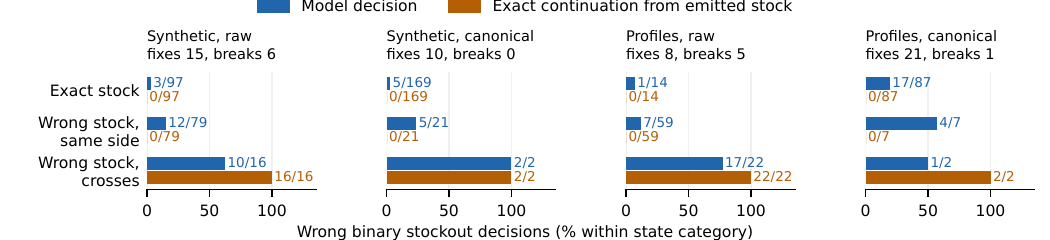}
  \caption{GPT-5.4 mini (low): wrong binary stockout decisions within each state category,
  for the model's decision and for exact continuation from its emitted stock (``fixes'':
  wrong model decisions made right by continuation; ``breaks'': right decisions made wrong).
  Continuation is error-free by construction in the first two rows and always wrong in the
  third; the empirical content is the category sizes and the model decisions. The same-side
  share depends on how far true stocks sit from the threshold: median distance 43 units on
  synthetic histories and 11 on profiles. Descriptive percentages.}
  \label{fig:decision}
\end{figure}
\input{diagnostic_findings}
\paragraph{An observed failure with a scenario consequence.}
As a post-hoc illustration, primary base \texttt{b0007}, mini raw repetition 0,
returns starting stock 31 instead of 18. Its future outputs are terminal stock 6,
zero loss, and no stockout. Exact rollout from its emitted stock of 31 produces
precisely those outputs, while rollout from the correct stock of 18 ends at zero,
loses seven units, and first stocks out at time 130. This trace is consistent
with a wrong-state rollout, without establishing the model's internal process.
It is illustrative rather than a separately sampled test. Both repetitions of this
request return the identical five fields, so repeat agreement would not have flagged
the error.

For any consistent rollout, terminal stock minus future loss equals starting
stock plus supplied receipts minus supplied demand. Consequently, the correct
terminal-stock/loss pair cannot be obtained by consistent replay from a wrong
starting stock in this environment. An output with wrong stock and correct
future quantities would demonstrate observable inconsistency, not reliable
recovery. Historical-loss errors can still coexist with correct starting stock
and future outputs.

\paragraph{Descriptive task-length strata.}
Mini raw all-five success is 39/64, 16/64, and 8/64 for 12, 36, and 72 historical
events, respectively; future-joint success is 43/64, 18/64, and 11/64. These are
different seeded histories within the factorial design, not matched
interventions identifying an isolated length or temporal-reasoning effect.
All configuration and factor strata remain available in the analysis artifact.

\paragraph{Why correct totals do not certify replay.}
A post-hoc snapshot trace (mini, primary \texttt{b0019}, repetition 0) copies
starting stock 9 correctly but emits final stock 0 and future loss 79 instead
of 8 and 87. Both pairs satisfy the same conservation equation. The final
receipt adds eight units after the last demand; lost sales cannot be filled
retroactively, so these eight units must remain. This shows why event-order
execution remains necessary after correct historical reconstruction, without
inferring the model's internal algorithm.

\paragraph{Executable continuation (post-hoc).}
\label{app:continuation}
Table~\ref{tab:continuationfull} retains all input arms and configurations for
the two main cohorts; the supplement also includes both control cohorts.
This is a deterministic transformation of original answers, not newly sampled
model behavior. The original prompt requests all five fields, and the strict
five-field validity gate is retained even if an otherwise invalid answer has a
usable stock field. For valid answers, continuation future-joint accuracy equals
starting-stock accuracy: conservation makes correct final stock and loss
impossible under consistent replay from an incorrect stock. This equivalence is
an endpoint property, not a new inventory algorithm. An accidentally correct
future triple paired with wrong stock can regress; this occurs once for mini's
raw retail responses. No such case is removed.

The operational false-alarm/miss endpoints need not move together. For mini's
canonical synthetic responses, execution reduces false alarms from 5/70 to 1/70
and misses from 7/122 to 1/122; canonical retail rates improve from 15/46 to 1/46
and 7/50 to 1/50. For raw retail responses, false alarms rise from 15/46 to 19/46
while misses fall from 10/49 to 3/49. These rates condition on valid answers;
the invalid raw retail response belongs to a positive-stockout task and remains
a failure in the unconditional accuracy. They measure detection, not timing.
\input{continuation_full_table}

\section{Extension studies: protocol summary and full results}
\label{app:extensions}
Both extension studies were specified in a written protocol after the original
outcomes were known and before any of their calls, with stated hypotheses, and are
reported in full whatever they show. They reuse the archived heldout histories, so
they add configurations and arms rather than independent histories. The effort sweeps
hold every prompt, task and schema fixed and change only the provider's requested
reasoning effort (GPT-5.4 mini at medium and high on all arms; GPT-5.4 at medium and high on
raw and canonical input, added by Amendment 2). The code-inside arms add the provider's
hosted code-execution tool to the two OpenAI configurations with automatic tool choice,
so the model decides whether to run code; the instructions gain one sentence saying the
tool is available. The state-only arms withhold the future schedule and
horizon, ask for the two historical integers, and apply the schedule in code; invalid
two-field answers fail and are not repaired. A same-period replication of the original
low-effort synthetic cells was added by amendment, before its calls, to rule out a
collection-date confound in the effort comparison. Every response was reparsed and
rescored independently of the runner, the planned grids were verified complete, task
payloads were checked byte-for-byte against the archived cohorts, and the conservation
identity was verified on every hybrid answer. Full per-cell counts, paired differences
with intervals, token usage and cost are in the supplement's extension audit;
Table~\ref{tab:allcells} lists every cell and Tables~\ref{tab:extpaired_heldout}
and~\ref{tab:extpaired_profiles} every paired difference. The anonymous supplement
contains all code, generators, prompts, raw provider responses, audits and the
reasoning-summary traces of diagnostic D1.

\input{extension_appendix}


\section{Amendment 4: capability ladder, interfaces and transport}
\label{app:ladder}
Protocol \texttt{protocol\_v4\_extensions.md} (frozen before its calls, with transport notes
appended before each rerun) adds interface-matched GPT-5.4 cells (E6), Claude Haiku~4.5 (E7)
and six open-weight models in a requested-effort by tool-choice cross (E8), plus a same-period
GPT-5.4 forced-call replication (E6b). Open-weight models are served through OpenRouter with
one pinned upstream per model and fallbacks disabled; the served provider is stored in every
record and checked by the audit. Rate limiting by third-party upstreams (HTTP 429) is retried
with bounded backoff before any completion exists, and requests that exhaust their attempts
are re-issued in documented completion passes; a model output is never re-sampled. Two
grids were interrupted and repeated in full before any accuracy was inspected, and the Qwen
provider withdrew the model before the forced-call profile stage, so those four cells are
absent.
\input{ladder_appendix}
\input{interaction_raw_table}
The original forced-call placement table and the mini/GPT-5.4 requested-effort table follow
for comparison with the interface-matched Table~\ref{tab:ladder}.
\input{extension_state_table}
\input{extension_effort_table}

%% file: coverage_text.tex
Oracle task coverage counts each base once, independently of model calls; these are scenario properties, not model success rates.
The synthetic cohort has 52/96 bases with positive historical loss, 61/96 with positive future loss, and 56/96 with positive terminal stock. These categories can overlap.
The external-profile cohort has 28/48 bases with positive historical loss, 25/48 with positive future loss, and 37/48 with positive terminal stock. These categories can overlap.

%% file: main_table.tex
\begin{table}[t]
\centering
\small
\setlength{\tabcolsep}{3pt}
\begin{tabular}{lrrr}
\toprule
Configuration & Raw & Canonical & Snapshot \\
\midrule
GPT-5.4 mini (low) & 32.8 / 37.5 & 61.5 / 76.0 & 89.6 / 89.6 \\
GPT-5.4 (low) & 83.9 / 88.0 & 94.8 / 98.4 & 99.0 / 99.0 \\
Sonnet 4.6 & 97.9 / 99.0 & 99.0 / 99.5 & 100.0 / 100.0 \\
Sonnet + thinking & 96.9 / 97.9 & 100.0 / 100.0 & 100.0 / 100.0 \\
\bottomrule
\end{tabular}
\caption{Held-out synthetic replay: each cell is all-five / future-joint exact accuracy (\%), 192 calls per configuration/protocol. The first endpoint requires all five outputs; the second requires all three future-scenario outputs. Invalid contracts fail. Thinking denotes a separate configuration.}
\label{tab:main}
\end{table}

%% file: external_table.tex
\begin{table}[t]
\centering
\small
\setlength{\tabcolsep}{3pt}
\begin{tabular}{lrrr}
\toprule
Configuration & Raw & Canonical & Snapshot \\
\midrule
GPT-5.4 mini (low) & 4.2 / 12.5 & 40.6 / 46.9 & 89.6 / 89.6 \\
GPT-5.4 (low) & 88.5 / 90.6 & 91.7 / 91.7 & 99.0 / 99.0 \\
Sonnet 4.6 & 91.7 / 95.8 & 97.9 / 99.0 & 100.0 / 100.0 \\
Sonnet + thinking & 96.9 / 97.9 & 100.0 / 100.0 & 100.0 / 100.0 \\
\bottomrule
\end{tabular}
\caption{External-profile simulation: each cell is all-five / future-joint exact accuracy (\%), 96 calls per configuration/protocol. The first endpoint requires all five outputs; the second requires all three future-scenario outputs. Invalid contracts fail. Thinking denotes a separate configuration.}
\label{tab:external}
\end{table}

%% file: factor_table.tex
\begin{table}[htbp]
\centering\small
\begin{tabular}{lrrrr}
\toprule
Stratum & \multicolumn{2}{c}{Stock exact} & \multicolumn{2}{c}{Future joint} \\
 & raw & canonical & raw & canonical \\
\midrule
immediate reporting & 63/96 & 86/96 & 46/96 & 71/96 \\
delayed reporting & 34/96 & 83/96 & 26/96 & 75/96 \\
no corrections & 62/96 & 91/96 & 46/96 & 76/96 \\
corrections present & 35/96 & 78/96 & 26/96 & 70/96 \\
12 events & 50/64 & 61/64 & 43/64 & 59/64 \\
36 events & 26/64 & 59/64 & 18/64 & 51/64 \\
72 events & 21/64 & 49/64 & 11/64 & 36/64 \\
\bottomrule
\end{tabular}
\caption{GPT-5.4 mini (low) on the synthetic factorial, by design factor (96 calls per stratum). Strata are different seeded histories, not matched interventions; counts are descriptive.}
\label{tab:factors}
\end{table}

%% file: extension_all_cells_table.tex
{\scriptsize\setlength{\tabcolsep}{2pt}
\begin{longtable}{lllrrrrrrr}
\caption{All cells of the original grid and the extension studies (\%). State-only arms report the hybrid answer (model historical fields plus executable continuation); their future-joint accuracy equals their stock accuracy by conservation. Tool arms had a code-execution tool available. Invalid answers (missing, malformed or negative) fail every endpoint and remain in denominators.}\label{tab:allcells} \\
\toprule
Cohort & Configuration & Arm & N & Invalid & Missing & All five & Stock & Hist. joint & Future joint \\
\midrule
\endfirsthead
\multicolumn{10}{l}{\tablename~\thetable{} (continued)} \\
\toprule
Cohort & Configuration & Arm & N & Invalid & Missing & All five & Stock & Hist. joint & Future joint \\
\midrule
\endhead
\midrule
\multicolumn{10}{r}{continued on next page} \\
\endfoot
\bottomrule
\endlastfoot
Synthetic & GPT-5.4 mini (low) & raw & 192 & 0 & 0 & 32.8 & 50.5 & 43.2 & 37.5 \\
Synthetic & GPT-5.4 mini (low) & canonical & 192 & 0 & 0 & 61.5 & 88.0 & 69.3 & 76.0 \\
Synthetic & GPT-5.4 mini (low) & snapshot & 192 & 0 & 0 & 89.6 & 100.0 & 100.0 & 89.6 \\
Synthetic & GPT-5.4 mini (low) & state-raw & 192 & 0 & 0 & 43.2 & 55.2 & 43.2 & 55.2 \\
Synthetic & GPT-5.4 mini (low) & state-canonical & 192 & 0 & 0 & 73.4 & 93.8 & 73.4 & 93.8 \\
Synthetic & GPT-5.4 mini (low) & tool-raw & 192 & 0 & 0 & 90.1 & 94.8 & 91.1 & 91.7 \\
Synthetic & GPT-5.4 mini (low) & tool-canonical & 192 & 0 & 0 & 94.8 & 99.5 & 95.8 & 97.9 \\
Synthetic & GPT-5.4 mini (medium) & raw & 192 & 2 & 0 & 23.4 & 45.8 & 35.4 & 26.6 \\
Synthetic & GPT-5.4 mini (medium) & canonical & 192 & 1 & 0 & 43.2 & 66.1 & 58.3 & 46.9 \\
Synthetic & GPT-5.4 mini (medium) & snapshot & 192 & 0 & 0 & 97.9 & 100.0 & 100.0 & 97.9 \\
Synthetic & GPT-5.4 mini (high) & raw & 192 & 0 & 0 & 22.4 & 41.7 & 34.4 & 25.0 \\
Synthetic & GPT-5.4 mini (high) & canonical & 192 & 1 & 0 & 33.9 & 56.2 & 52.1 & 34.9 \\
Synthetic & GPT-5.4 mini (high) & snapshot & 192 & 0 & 0 & 97.4 & 100.0 & 100.0 & 97.4 \\
Synthetic & GPT-5.4 mini (high) & state-raw & 192 & 0 & 0 & 45.3 & 54.2 & 45.3 & 54.2 \\
Synthetic & GPT-5.4 mini (high) & state-canonical & 192 & 1 & 0 & 64.6 & 71.4 & 64.6 & 71.4 \\
Synthetic & GPT-5.4 mini (low, auto tool choice) & raw & 192 & 1 & 1 & 46.4 & 58.9 & 47.9 & 56.8 \\
Synthetic & GPT-5.4 mini (low, auto tool choice) & canonical & 192 & 0 & 0 & 72.4 & 91.7 & 76.6 & 86.5 \\
Synthetic & GPT-5.4 mini (low, auto tool choice) & snapshot & 192 & 0 & 0 & 94.8 & 100.0 & 100.0 & 94.8 \\
Synthetic & GPT-5.4 mini (low, auto tool choice) & state-raw & 192 & 0 & 0 & 53.6 & 69.3 & 53.6 & 69.3 \\
Synthetic & GPT-5.4 mini (low, auto tool choice) & state-canonical & 192 & 0 & 0 & 80.2 & 94.8 & 80.2 & 94.8 \\
Synthetic & GPT-5.4 mini (high, auto tool choice) & raw & 192 & 1 & 1 & 72.4 & 82.3 & 75.5 & 79.2 \\
Synthetic & GPT-5.4 mini (high, auto tool choice) & canonical & 192 & 0 & 0 & 89.1 & 96.9 & 92.7 & 93.2 \\
Synthetic & GPT-5.4 mini (high, auto tool choice) & snapshot & 192 & 0 & 0 & 98.4 & 100.0 & 100.0 & 98.4 \\
Synthetic & GPT-5.4 (low) & raw & 192 & 0 & 0 & 83.9 & 91.1 & 86.5 & 88.0 \\
Synthetic & GPT-5.4 (low) & canonical & 192 & 0 & 0 & 94.8 & 99.5 & 95.8 & 98.4 \\
Synthetic & GPT-5.4 (low) & snapshot & 192 & 0 & 0 & 99.0 & 100.0 & 100.0 & 99.0 \\
Synthetic & GPT-5.4 (low) & state-raw & 192 & 0 & 0 & 88.5 & 91.7 & 88.5 & 91.7 \\
Synthetic & GPT-5.4 (low) & state-canonical & 192 & 0 & 0 & 96.9 & 98.4 & 96.9 & 98.4 \\
Synthetic & GPT-5.4 (low) & tool-raw & 192 & 0 & 0 & 100.0 & 100.0 & 100.0 & 100.0 \\
Synthetic & GPT-5.4 (low) & tool-canonical & 192 & 0 & 0 & 96.9 & 98.4 & 97.9 & 97.4 \\
Synthetic & GPT-5.4 (medium) & raw & 192 & 0 & 0 & 95.3 & 98.4 & 98.4 & 95.3 \\
Synthetic & GPT-5.4 (medium) & canonical & 192 & 0 & 0 & 93.8 & 97.9 & 95.3 & 95.8 \\
Synthetic & GPT-5.4 (high) & raw & 192 & 0 & 0 & 94.3 & 96.9 & 96.4 & 94.3 \\
Synthetic & GPT-5.4 (high) & canonical & 192 & 0 & 0 & 94.3 & 98.4 & 97.9 & 94.8 \\
Synthetic & GPT-5.4 (low, auto tool choice) & raw & 192 & 0 & 0 & 96.4 & 98.4 & 96.4 & 98.4 \\
Synthetic & GPT-5.4 (low, auto tool choice) & canonical & 192 & 0 & 0 & 95.8 & 98.4 & 95.8 & 98.4 \\
Synthetic & GPT-5.4 (low, auto tool choice) & snapshot & 192 & 0 & 0 & 91.1 & 100.0 & 100.0 & 91.1 \\
Synthetic & GPT-5.4 (low, auto tool choice) & state-raw & 192 & 0 & 0 & 96.4 & 96.9 & 96.4 & 96.9 \\
Synthetic & GPT-5.4 (low, auto tool choice) & state-canonical & 192 & 0 & 0 & 81.2 & 83.9 & 81.2 & 83.9 \\
Synthetic & GPT-5.4 (high, auto tool choice) & raw & 192 & 0 & 0 & 99.0 & 99.5 & 99.5 & 99.0 \\
Synthetic & GPT-5.4 (high, auto tool choice) & canonical & 192 & 0 & 0 & 99.5 & 100.0 & 99.5 & 100.0 \\
Synthetic & Haiku 4.5 & raw & 192 & 0 & 0 & 43.8 & 59.4 & 47.9 & 52.6 \\
Synthetic & Haiku 4.5 & canonical & 192 & 0 & 0 & 79.2 & 96.9 & 87.0 & 85.9 \\
Synthetic & Haiku 4.5 & snapshot & 192 & 0 & 0 & 92.7 & 100.0 & 100.0 & 92.7 \\
Synthetic & Haiku 4.5 & state-raw & 192 & 0 & 0 & 44.3 & 61.5 & 44.3 & 61.5 \\
Synthetic & Haiku 4.5 & state-canonical & 192 & 0 & 0 & 86.5 & 92.7 & 86.5 & 92.7 \\
Synthetic & Haiku 4.5 + thinking & raw & 192 & 10 & 10 & 82.3 & 90.1 & 83.3 & 88.5 \\
Synthetic & Haiku 4.5 + thinking & canonical & 192 & 0 & 0 & 94.8 & 100.0 & 96.4 & 96.9 \\
Synthetic & Haiku 4.5 + thinking & snapshot & 192 & 0 & 0 & 99.0 & 100.0 & 100.0 & 99.0 \\
Synthetic & Haiku 4.5 + thinking & state-raw & 192 & 11 & 11 & 84.9 & 88.0 & 84.9 & 88.0 \\
Synthetic & Haiku 4.5 + thinking & state-canonical & 192 & 0 & 0 & 99.0 & 100.0 & 99.0 & 100.0 \\
Synthetic & Sonnet 4.6 & raw & 192 & 0 & 0 & 97.9 & 99.5 & 98.4 & 99.0 \\
Synthetic & Sonnet 4.6 & canonical & 192 & 0 & 0 & 99.0 & 100.0 & 99.5 & 99.5 \\
Synthetic & Sonnet 4.6 & snapshot & 192 & 0 & 0 & 100.0 & 100.0 & 100.0 & 100.0 \\
Synthetic & Sonnet 4.6 & state-raw & 192 & 0 & 0 & 99.5 & 100.0 & 99.5 & 100.0 \\
Synthetic & Sonnet 4.6 & state-canonical & 192 & 0 & 0 & 99.0 & 100.0 & 99.0 & 100.0 \\
Synthetic & Sonnet + thinking & raw & 192 & 0 & 0 & 96.9 & 99.0 & 97.9 & 97.9 \\
Synthetic & Sonnet + thinking & canonical & 192 & 0 & 0 & 100.0 & 100.0 & 100.0 & 100.0 \\
Synthetic & Sonnet + thinking & snapshot & 192 & 0 & 0 & 100.0 & 100.0 & 100.0 & 100.0 \\
Synthetic & Sonnet + thinking & state-raw & 192 & 0 & 0 & 99.0 & 100.0 & 99.0 & 100.0 \\
Synthetic & Sonnet + thinking & state-canonical & 192 & 0 & 0 & 99.5 & 100.0 & 99.5 & 100.0 \\
Synthetic & gpt-oss-20b (low, forced call) & raw & 192 & 85 & 85 & 20.3 & 31.2 & 26.6 & 23.4 \\
Synthetic & gpt-oss-20b (low, forced call) & canonical & 192 & 59 & 59 & 50.5 & 58.3 & 54.7 & 52.1 \\
Synthetic & gpt-oss-20b (low, auto tool choice) & raw & 192 & 88 & 88 & 20.8 & 31.8 & 25.5 & 24.0 \\
Synthetic & gpt-oss-20b (low, auto tool choice) & canonical & 192 & 51 & 51 & 46.4 & 62.5 & 52.1 & 51.0 \\
Synthetic & gpt-oss-20b (low, auto tool choice) & snapshot & 192 & 0 & 0 & 82.8 & 100.0 & 100.0 & 82.8 \\
Synthetic & gpt-oss-20b (low, auto tool choice) & state-raw & 192 & 53 & 53 & 29.7 & 37.5 & 29.7 & 37.5 \\
Synthetic & gpt-oss-20b (low, auto tool choice) & state-canonical & 192 & 32 & 32 & 57.8 & 71.4 & 57.8 & 71.4 \\
Synthetic & gpt-oss-20b (high, forced call) & raw & 192 & 106 & 106 & 42.2 & 42.2 & 42.2 & 42.2 \\
Synthetic & gpt-oss-20b (high, forced call) & canonical & 192 & 65 & 65 & 66.1 & 66.1 & 66.1 & 66.1 \\
Synthetic & gpt-oss-20b (high, auto tool choice) & raw & 192 & 112 & 112 & 39.1 & 40.6 & 40.6 & 39.1 \\
Synthetic & gpt-oss-20b (high, auto tool choice) & canonical & 192 & 75 & 75 & 59.4 & 60.4 & 60.4 & 59.4 \\
Synthetic & gpt-oss-120b (low, forced call) & raw & 192 & 0 & 0 & 42.7 & 53.6 & 46.4 & 45.8 \\
Synthetic & gpt-oss-120b (low, forced call) & canonical & 192 & 0 & 0 & 62.0 & 79.7 & 68.8 & 69.3 \\
Synthetic & gpt-oss-120b (low, auto tool choice) & raw & 192 & 41 & 41 & 46.9 & 54.7 & 50.5 & 49.5 \\
Synthetic & gpt-oss-120b (low, auto tool choice) & canonical & 192 & 12 & 12 & 67.7 & 78.1 & 69.8 & 73.4 \\
Synthetic & gpt-oss-120b (low, auto tool choice) & snapshot & 192 & 1 & 1 & 94.8 & 99.0 & 99.0 & 94.8 \\
Synthetic & gpt-oss-120b (low, auto tool choice) & state-raw & 192 & 16 & 16 & 48.4 & 57.8 & 48.4 & 57.8 \\
Synthetic & gpt-oss-120b (low, auto tool choice) & state-canonical & 192 & 4 & 4 & 68.2 & 76.0 & 68.2 & 76.0 \\
Synthetic & gpt-oss-120b (high, forced call) & raw & 192 & 89 & 89 & 53.6 & 53.6 & 53.6 & 53.6 \\
Synthetic & gpt-oss-120b (high, forced call) & canonical & 192 & 65 & 65 & 66.1 & 66.1 & 66.1 & 66.1 \\
Synthetic & gpt-oss-120b (high, auto tool choice) & raw & 192 & 98 & 98 & 49.0 & 49.0 & 49.0 & 49.0 \\
Synthetic & gpt-oss-120b (high, auto tool choice) & canonical & 192 & 60 & 60 & 68.8 & 68.8 & 68.8 & 68.8 \\
Synthetic & Qwen3.8-27B (low, forced call) & raw & 192 & 41 & 40 & 75.5 & 77.1 & 76.0 & 76.6 \\
Synthetic & Qwen3.8-27B (low, forced call) & canonical & 192 & 16 & 16 & 91.7 & 91.7 & 91.7 & 91.7 \\
Synthetic & Qwen3.8-27B (low, auto tool choice) & raw & 192 & 3 & 3 & 92.2 & 95.3 & 92.2 & 95.3 \\
Synthetic & Qwen3.8-27B (low, auto tool choice) & canonical & 192 & 0 & 0 & 98.4 & 99.5 & 98.4 & 99.5 \\
Synthetic & Qwen3.8-27B (low, auto tool choice) & snapshot & 192 & 0 & 0 & 100.0 & 100.0 & 100.0 & 100.0 \\
Synthetic & Qwen3.8-27B (low, auto tool choice) & state-raw & 192 & 3 & 3 & 95.8 & 96.9 & 95.8 & 96.9 \\
Synthetic & Qwen3.8-27B (low, auto tool choice) & state-canonical & 192 & 0 & 0 & 99.5 & 100.0 & 99.5 & 100.0 \\
Synthetic & Qwen3.8-27B (high, forced call) & raw & 192 & 28 & 28 & 77.1 & 81.8 & 78.1 & 80.2 \\
Synthetic & Qwen3.8-27B (high, forced call) & canonical & 192 & 10 & 10 & 90.6 & 94.3 & 93.2 & 91.7 \\
Synthetic & Qwen3.8-27B (high, auto tool choice) & raw & 192 & 19 & 19 & 85.9 & 88.5 & 85.9 & 88.5 \\
Synthetic & Qwen3.8-27B (high, auto tool choice) & canonical & 192 & 8 & 8 & 91.1 & 95.8 & 94.3 & 92.7 \\
Synthetic & DeepSeek V4 Flash (low, forced call) & raw & 192 & 8 & 4 & 2.1 & 6.2 & 2.1 & 2.1 \\
Synthetic & DeepSeek V4 Flash (low, forced call) & canonical & 192 & 10 & 6 & 2.1 & 7.8 & 2.6 & 2.1 \\
Synthetic & DeepSeek V4 Flash (low, auto tool choice) & raw & 192 & 0 & 0 & 94.3 & 98.4 & 94.3 & 98.4 \\
Synthetic & DeepSeek V4 Flash (low, auto tool choice) & canonical & 192 & 0 & 0 & 97.9 & 99.0 & 97.9 & 99.0 \\
Synthetic & DeepSeek V4 Flash (low, auto tool choice) & snapshot & 192 & 0 & 0 & 99.0 & 100.0 & 100.0 & 99.0 \\
Synthetic & DeepSeek V4 Flash (low, auto tool choice) & state-raw & 192 & 0 & 0 & 94.8 & 99.5 & 94.8 & 99.5 \\
Synthetic & DeepSeek V4 Flash (low, auto tool choice) & state-canonical & 192 & 1 & 0 & 97.4 & 99.0 & 97.4 & 99.0 \\
Synthetic & DeepSeek V4 Flash (high, forced call) & raw & 192 & 12 & 10 & 1.6 & 6.8 & 1.6 & 1.6 \\
Synthetic & DeepSeek V4 Flash (high, forced call) & canonical & 192 & 18 & 15 & 1.0 & 3.6 & 1.0 & 1.0 \\
Synthetic & DeepSeek V4 Flash (high, auto tool choice) & raw & 192 & 2 & 2 & 92.2 & 94.3 & 93.2 & 93.2 \\
Synthetic & DeepSeek V4 Flash (high, auto tool choice) & canonical & 192 & 0 & 0 & 96.4 & 97.9 & 96.9 & 97.4 \\
Synthetic & GLM-5.3 Flash (low, forced call) & raw & 192 & 0 & 0 & 38.0 & 78.1 & 57.8 & 52.6 \\
Synthetic & GLM-5.3 Flash (low, forced call) & canonical & 192 & 0 & 0 & 53.6 & 84.4 & 69.8 & 62.5 \\
Synthetic & GLM-5.3 Flash (low, auto tool choice) & raw & 192 & 0 & 0 & 43.8 & 76.0 & 59.4 & 55.7 \\
Synthetic & GLM-5.3 Flash (low, auto tool choice) & canonical & 192 & 2 & 2 & 50.5 & 85.9 & 67.7 & 64.1 \\
Synthetic & GLM-5.3 Flash (low, auto tool choice) & snapshot & 192 & 0 & 0 & 58.9 & 100.0 & 100.0 & 58.9 \\
Synthetic & GLM-5.3 Flash (low, auto tool choice) & state-raw & 192 & 0 & 0 & 52.6 & 77.6 & 52.6 & 77.6 \\
Synthetic & GLM-5.3 Flash (low, auto tool choice) & state-canonical & 192 & 0 & 0 & 69.3 & 87.0 & 69.3 & 87.0 \\
Synthetic & GLM-5.3 Flash (high, forced call) & raw & 192 & 0 & 0 & 62.5 & 91.1 & 74.5 & 74.5 \\
Synthetic & GLM-5.3 Flash (high, forced call) & canonical & 192 & 0 & 0 & 74.5 & 93.8 & 83.3 & 82.3 \\
Synthetic & GLM-5.3 Flash (high, auto tool choice) & raw & 192 & 1 & 1 & 66.7 & 92.2 & 79.2 & 76.6 \\
Synthetic & GLM-5.3 Flash (high, auto tool choice) & canonical & 192 & 0 & 0 & 74.5 & 95.8 & 83.3 & 84.4 \\
Synthetic & DeepSeek V4 Pro (low, forced call) & raw & 192 & 6 & 6 & 94.3 & 95.8 & 94.3 & 96.4 \\
Synthetic & DeepSeek V4 Pro (low, forced call) & canonical & 192 & 2 & 2 & 97.4 & 99.0 & 97.9 & 98.4 \\
Synthetic & DeepSeek V4 Pro (low, auto tool choice) & raw & 192 & 0 & 0 & 98.4 & 99.0 & 98.4 & 99.0 \\
Synthetic & DeepSeek V4 Pro (low, auto tool choice) & canonical & 192 & 0 & 0 & 97.4 & 100.0 & 98.4 & 99.0 \\
Synthetic & DeepSeek V4 Pro (low, auto tool choice) & snapshot & 192 & 0 & 0 & 99.0 & 100.0 & 100.0 & 99.0 \\
Synthetic & DeepSeek V4 Pro (low, auto tool choice) & state-raw & 192 & 2 & 2 & 97.9 & 98.4 & 97.9 & 98.4 \\
Synthetic & DeepSeek V4 Pro (low, auto tool choice) & state-canonical & 192 & 0 & 0 & 96.9 & 99.0 & 96.9 & 99.0 \\
Synthetic & DeepSeek V4 Pro (high, forced call) & raw & 192 & 16 & 16 & 91.7 & 91.7 & 91.7 & 91.7 \\
Synthetic & DeepSeek V4 Pro (high, forced call) & canonical & 192 & 32 & 32 & 80.2 & 83.3 & 81.2 & 82.3 \\
Synthetic & DeepSeek V4 Pro (high, auto tool choice) & raw & 192 & 12 & 12 & 92.7 & 93.8 & 93.2 & 93.2 \\
Synthetic & DeepSeek V4 Pro (high, auto tool choice) & canonical & 192 & 2 & 2 & 96.4 & 97.4 & 96.4 & 97.4 \\
\midrule
Retail profiles & GPT-5.4 mini (low) & raw & 96 & 1 & 0 & 4.2 & 14.6 & 4.2 & 12.5 \\
Retail profiles & GPT-5.4 mini (low) & canonical & 96 & 0 & 0 & 40.6 & 90.6 & 74.0 & 46.9 \\
Retail profiles & GPT-5.4 mini (low) & snapshot & 96 & 0 & 0 & 89.6 & 100.0 & 100.0 & 89.6 \\
Retail profiles & GPT-5.4 mini (low) & state-raw & 96 & 0 & 0 & 3.1 & 17.7 & 3.1 & 17.7 \\
Retail profiles & GPT-5.4 mini (low) & state-canonical & 96 & 0 & 0 & 89.6 & 99.0 & 89.6 & 99.0 \\
Retail profiles & GPT-5.4 mini (low) & tool-raw & 96 & 0 & 0 & 87.5 & 89.6 & 87.5 & 89.6 \\
Retail profiles & GPT-5.4 mini (low) & tool-canonical & 96 & 0 & 0 & 97.9 & 100.0 & 97.9 & 100.0 \\
Retail profiles & GPT-5.4 mini (medium) & raw & 96 & 0 & 0 & 11.5 & 24.0 & 11.5 & 16.7 \\
Retail profiles & GPT-5.4 mini (medium) & canonical & 96 & 0 & 0 & 36.5 & 52.1 & 41.7 & 38.5 \\
Retail profiles & GPT-5.4 mini (medium) & snapshot & 96 & 0 & 0 & 94.8 & 100.0 & 100.0 & 94.8 \\
Retail profiles & GPT-5.4 mini (high) & raw & 96 & 1 & 0 & 13.5 & 30.2 & 13.5 & 20.8 \\
Retail profiles & GPT-5.4 mini (high) & canonical & 96 & 0 & 0 & 27.1 & 35.4 & 28.1 & 29.2 \\
Retail profiles & GPT-5.4 mini (high) & snapshot & 96 & 0 & 0 & 93.8 & 100.0 & 100.0 & 93.8 \\
Retail profiles & GPT-5.4 mini (high) & state-raw & 96 & 0 & 0 & 21.9 & 34.4 & 21.9 & 34.4 \\
Retail profiles & GPT-5.4 mini (high) & state-canonical & 96 & 0 & 0 & 26.0 & 34.4 & 26.0 & 34.4 \\
Retail profiles & GPT-5.4 mini (low, auto tool choice) & raw & 96 & 0 & 0 & 4.2 & 18.8 & 5.2 & 14.6 \\
Retail profiles & GPT-5.4 mini (low, auto tool choice) & canonical & 96 & 1 & 1 & 71.9 & 92.7 & 79.2 & 82.3 \\
Retail profiles & GPT-5.4 mini (low, auto tool choice) & snapshot & 96 & 0 & 0 & 89.6 & 100.0 & 100.0 & 89.6 \\
Retail profiles & GPT-5.4 mini (low, auto tool choice) & state-raw & 96 & 1 & 1 & 4.2 & 19.8 & 4.2 & 19.8 \\
Retail profiles & GPT-5.4 mini (low, auto tool choice) & state-canonical & 96 & 2 & 2 & 84.4 & 94.8 & 84.4 & 94.8 \\
Retail profiles & GPT-5.4 mini (high, auto tool choice) & raw & 96 & 1 & 1 & 35.4 & 49.0 & 35.4 & 49.0 \\
Retail profiles & GPT-5.4 mini (high, auto tool choice) & canonical & 96 & 1 & 1 & 93.8 & 96.9 & 95.8 & 94.8 \\
Retail profiles & GPT-5.4 (low) & raw & 96 & 0 & 0 & 88.5 & 90.6 & 88.5 & 90.6 \\
Retail profiles & GPT-5.4 (low) & canonical & 96 & 0 & 0 & 91.7 & 100.0 & 100.0 & 91.7 \\
Retail profiles & GPT-5.4 (low) & snapshot & 96 & 0 & 0 & 99.0 & 100.0 & 100.0 & 99.0 \\
Retail profiles & GPT-5.4 (low) & state-raw & 96 & 0 & 0 & 85.4 & 90.6 & 85.4 & 90.6 \\
Retail profiles & GPT-5.4 (low) & state-canonical & 96 & 0 & 0 & 99.0 & 99.0 & 99.0 & 99.0 \\
Retail profiles & GPT-5.4 (low) & tool-raw & 96 & 0 & 0 & 100.0 & 100.0 & 100.0 & 100.0 \\
Retail profiles & GPT-5.4 (low) & tool-canonical & 96 & 0 & 0 & 100.0 & 100.0 & 100.0 & 100.0 \\
Retail profiles & GPT-5.4 (medium) & raw & 96 & 0 & 0 & 95.8 & 95.8 & 95.8 & 95.8 \\
Retail profiles & GPT-5.4 (medium) & canonical & 96 & 0 & 0 & 94.8 & 99.0 & 99.0 & 94.8 \\
Retail profiles & GPT-5.4 (high) & raw & 96 & 0 & 0 & 99.0 & 99.0 & 99.0 & 99.0 \\
Retail profiles & GPT-5.4 (high) & canonical & 96 & 0 & 0 & 88.5 & 90.6 & 90.6 & 88.5 \\
Retail profiles & GPT-5.4 (low, auto tool choice) & raw & 96 & 0 & 0 & 95.8 & 97.9 & 95.8 & 97.9 \\
Retail profiles & GPT-5.4 (low, auto tool choice) & canonical & 96 & 0 & 0 & 100.0 & 100.0 & 100.0 & 100.0 \\
Retail profiles & GPT-5.4 (low, auto tool choice) & snapshot & 96 & 0 & 0 & 89.6 & 100.0 & 100.0 & 89.6 \\
Retail profiles & GPT-5.4 (low, auto tool choice) & state-raw & 96 & 0 & 0 & 93.8 & 95.8 & 93.8 & 95.8 \\
Retail profiles & GPT-5.4 (low, auto tool choice) & state-canonical & 96 & 0 & 0 & 91.7 & 91.7 & 91.7 & 91.7 \\
Retail profiles & GPT-5.4 (high, auto tool choice) & raw & 96 & 0 & 0 & 100.0 & 100.0 & 100.0 & 100.0 \\
Retail profiles & GPT-5.4 (high, auto tool choice) & canonical & 96 & 0 & 0 & 100.0 & 100.0 & 100.0 & 100.0 \\
Retail profiles & Haiku 4.5 & raw & 96 & 0 & 0 & 1.0 & 7.3 & 1.0 & 6.2 \\
Retail profiles & Haiku 4.5 & canonical & 96 & 0 & 0 & 71.9 & 96.9 & 77.1 & 89.6 \\
Retail profiles & Haiku 4.5 & snapshot & 96 & 0 & 0 & 92.7 & 100.0 & 100.0 & 92.7 \\
Retail profiles & Haiku 4.5 & state-raw & 96 & 0 & 0 & 1.0 & 6.2 & 1.0 & 6.2 \\
Retail profiles & Haiku 4.5 & state-canonical & 96 & 0 & 0 & 78.1 & 94.8 & 78.1 & 94.8 \\
Retail profiles & Haiku 4.5 + thinking & raw & 96 & 37 & 37 & 33.3 & 41.7 & 34.4 & 40.6 \\
Retail profiles & Haiku 4.5 + thinking & canonical & 96 & 0 & 0 & 87.5 & 99.0 & 93.8 & 91.7 \\
Retail profiles & Haiku 4.5 + thinking & snapshot & 96 & 0 & 0 & 96.9 & 100.0 & 100.0 & 96.9 \\
Retail profiles & Haiku 4.5 + thinking & state-raw & 96 & 22 & 21 & 46.9 & 59.4 & 46.9 & 59.4 \\
Retail profiles & Haiku 4.5 + thinking & state-canonical & 96 & 0 & 0 & 94.8 & 99.0 & 94.8 & 99.0 \\
Retail profiles & Sonnet 4.6 & raw & 96 & 0 & 0 & 91.7 & 96.9 & 91.7 & 95.8 \\
Retail profiles & Sonnet 4.6 & canonical & 96 & 0 & 0 & 97.9 & 100.0 & 99.0 & 99.0 \\
Retail profiles & Sonnet 4.6 & snapshot & 96 & 0 & 0 & 100.0 & 100.0 & 100.0 & 100.0 \\
Retail profiles & Sonnet 4.6 & state-raw & 96 & 0 & 0 & 92.7 & 95.8 & 92.7 & 95.8 \\
Retail profiles & Sonnet 4.6 & state-canonical & 96 & 0 & 0 & 99.0 & 100.0 & 99.0 & 100.0 \\
Retail profiles & Sonnet + thinking & raw & 96 & 0 & 0 & 96.9 & 99.0 & 97.9 & 97.9 \\
Retail profiles & Sonnet + thinking & canonical & 96 & 0 & 0 & 100.0 & 100.0 & 100.0 & 100.0 \\
Retail profiles & Sonnet + thinking & snapshot & 96 & 0 & 0 & 100.0 & 100.0 & 100.0 & 100.0 \\
Retail profiles & Sonnet + thinking & state-raw & 96 & 0 & 0 & 96.9 & 100.0 & 96.9 & 100.0 \\
Retail profiles & Sonnet + thinking & state-canonical & 96 & 0 & 0 & 100.0 & 100.0 & 100.0 & 100.0 \\
Retail profiles & gpt-oss-20b (low, forced call) & raw & 96 & 73 & 73 & 0.0 & 3.1 & 0.0 & 1.0 \\
Retail profiles & gpt-oss-20b (low, forced call) & canonical & 96 & 53 & 53 & 7.3 & 17.7 & 9.4 & 11.5 \\
Retail profiles & gpt-oss-20b (low, auto tool choice) & raw & 96 & 69 & 69 & 0.0 & 2.1 & 0.0 & 0.0 \\
Retail profiles & gpt-oss-20b (low, auto tool choice) & canonical & 96 & 58 & 58 & 14.6 & 19.8 & 16.7 & 17.7 \\
Retail profiles & gpt-oss-20b (low, auto tool choice) & snapshot & 96 & 1 & 1 & 77.1 & 97.9 & 97.9 & 77.1 \\
Retail profiles & gpt-oss-20b (high, forced call) & raw & 96 & 96 & 96 & 0.0 & 0.0 & 0.0 & 0.0 \\
Retail profiles & gpt-oss-20b (high, forced call) & canonical & 96 & 49 & 49 & 45.8 & 47.9 & 45.8 & 47.9 \\
Retail profiles & gpt-oss-20b (high, auto tool choice) & raw & 96 & 96 & 96 & 0.0 & 0.0 & 0.0 & 0.0 \\
Retail profiles & gpt-oss-20b (high, auto tool choice) & canonical & 96 & 50 & 50 & 46.9 & 47.9 & 47.9 & 46.9 \\
Retail profiles & gpt-oss-120b (low, forced call) & raw & 96 & 0 & 0 & 1.0 & 17.7 & 3.1 & 7.3 \\
Retail profiles & gpt-oss-120b (low, forced call) & canonical & 96 & 0 & 0 & 69.8 & 85.4 & 78.1 & 76.0 \\
Retail profiles & gpt-oss-120b (low, auto tool choice) & raw & 96 & 47 & 47 & 2.1 & 10.4 & 2.1 & 5.2 \\
Retail profiles & gpt-oss-120b (low, auto tool choice) & canonical & 96 & 1 & 1 & 70.8 & 83.3 & 79.2 & 72.9 \\
Retail profiles & gpt-oss-120b (low, auto tool choice) & snapshot & 96 & 1 & 0 & 86.5 & 97.9 & 97.9 & 86.5 \\
Retail profiles & gpt-oss-120b (high, forced call) & raw & 96 & 96 & 96 & 0.0 & 0.0 & 0.0 & 0.0 \\
Retail profiles & gpt-oss-120b (high, forced call) & canonical & 96 & 85 & 85 & 11.5 & 11.5 & 11.5 & 11.5 \\
Retail profiles & gpt-oss-120b (high, auto tool choice) & raw & 96 & 96 & 96 & 0.0 & 0.0 & 0.0 & 0.0 \\
Retail profiles & gpt-oss-120b (high, auto tool choice) & canonical & 96 & 86 & 86 & 10.4 & 10.4 & 10.4 & 10.4 \\
Retail profiles & Qwen3.8-27B (low, auto tool choice) & raw & 96 & 0 & 0 & 93.8 & 96.9 & 93.8 & 96.9 \\
Retail profiles & Qwen3.8-27B (low, auto tool choice) & canonical & 96 & 0 & 0 & 99.0 & 100.0 & 100.0 & 99.0 \\
Retail profiles & Qwen3.8-27B (low, auto tool choice) & snapshot & 96 & 0 & 0 & 100.0 & 100.0 & 100.0 & 100.0 \\
Retail profiles & Qwen3.8-27B (high, auto tool choice) & raw & 96 & 0 & 0 & 96.9 & 97.9 & 96.9 & 97.9 \\
Retail profiles & Qwen3.8-27B (high, auto tool choice) & canonical & 96 & 0 & 0 & 95.8 & 99.0 & 97.9 & 96.9 \\
Retail profiles & DeepSeek V4 Flash (low, forced call) & raw & 96 & 0 & 0 & 0.0 & 7.3 & 1.0 & 0.0 \\
Retail profiles & DeepSeek V4 Flash (low, forced call) & canonical & 96 & 7 & 4 & 0.0 & 10.4 & 0.0 & 1.0 \\
Retail profiles & DeepSeek V4 Flash (low, auto tool choice) & raw & 96 & 2 & 1 & 91.7 & 94.8 & 91.7 & 94.8 \\
Retail profiles & DeepSeek V4 Flash (low, auto tool choice) & canonical & 96 & 1 & 1 & 95.8 & 97.9 & 96.9 & 96.9 \\
Retail profiles & DeepSeek V4 Flash (low, auto tool choice) & snapshot & 96 & 0 & 0 & 99.0 & 100.0 & 100.0 & 99.0 \\
Retail profiles & DeepSeek V4 Flash (high, forced call) & raw & 96 & 11 & 8 & 0.0 & 5.2 & 2.1 & 1.0 \\
Retail profiles & DeepSeek V4 Flash (high, forced call) & canonical & 96 & 1 & 1 & 0.0 & 10.4 & 0.0 & 1.0 \\
Retail profiles & DeepSeek V4 Flash (high, auto tool choice) & raw & 96 & 3 & 3 & 93.8 & 94.8 & 93.8 & 94.8 \\
Retail profiles & DeepSeek V4 Flash (high, auto tool choice) & canonical & 96 & 1 & 0 & 99.0 & 99.0 & 99.0 & 99.0 \\
Retail profiles & GLM-5.3 Flash (low, forced call) & raw & 96 & 0 & 0 & 22.9 & 53.1 & 28.1 & 31.2 \\
Retail profiles & GLM-5.3 Flash (low, forced call) & canonical & 96 & 0 & 0 & 49.0 & 90.6 & 67.7 & 64.6 \\
Retail profiles & GLM-5.3 Flash (low, auto tool choice) & raw & 96 & 0 & 0 & 11.5 & 51.0 & 21.9 & 26.0 \\
Retail profiles & GLM-5.3 Flash (low, auto tool choice) & canonical & 96 & 0 & 0 & 43.8 & 87.5 & 66.7 & 58.3 \\
Retail profiles & GLM-5.3 Flash (low, auto tool choice) & snapshot & 96 & 0 & 0 & 38.5 & 100.0 & 100.0 & 38.5 \\
Retail profiles & GLM-5.3 Flash (high, forced call) & raw & 96 & 0 & 0 & 59.4 & 87.5 & 71.9 & 70.8 \\
Retail profiles & GLM-5.3 Flash (high, forced call) & canonical & 96 & 0 & 0 & 66.7 & 94.8 & 75.0 & 83.3 \\
Retail profiles & GLM-5.3 Flash (high, auto tool choice) & raw & 96 & 0 & 0 & 64.6 & 84.4 & 75.0 & 70.8 \\
Retail profiles & GLM-5.3 Flash (high, auto tool choice) & canonical & 96 & 0 & 0 & 68.8 & 94.8 & 76.0 & 84.4 \\
Retail profiles & DeepSeek V4 Pro (low, forced call) & raw & 96 & 7 & 7 & 90.6 & 91.7 & 90.6 & 91.7 \\
Retail profiles & DeepSeek V4 Pro (low, forced call) & canonical & 96 & 2 & 2 & 95.8 & 97.9 & 96.9 & 96.9 \\
Retail profiles & DeepSeek V4 Pro (low, auto tool choice) & raw & 96 & 0 & 0 & 94.8 & 97.9 & 96.9 & 95.8 \\
Retail profiles & DeepSeek V4 Pro (low, auto tool choice) & canonical & 96 & 0 & 0 & 99.0 & 100.0 & 99.0 & 100.0 \\
Retail profiles & DeepSeek V4 Pro (low, auto tool choice) & snapshot & 96 & 0 & 0 & 95.8 & 100.0 & 100.0 & 95.8 \\
Retail profiles & DeepSeek V4 Pro (high, forced call) & raw & 96 & 19 & 19 & 79.2 & 79.2 & 79.2 & 79.2 \\
Retail profiles & DeepSeek V4 Pro (high, forced call) & canonical & 96 & 20 & 20 & 78.1 & 78.1 & 78.1 & 78.1 \\
Retail profiles & DeepSeek V4 Pro (high, auto tool choice) & raw & 96 & 3 & 3 & 95.8 & 96.9 & 95.8 & 96.9 \\
Retail profiles & DeepSeek V4 Pro (high, auto tool choice) & canonical & 96 & 0 & 0 & 99.0 & 100.0 & 99.0 & 100.0 \\
\end{longtable}}

%% file: controls_table.tex
\begin{table}[t]
\centering
\small
\setlength{\tabcolsep}{3pt}
\begin{tabular}{lrrrr}
\toprule
Configuration & K-chg & K-both & D-same & D-both \\
\midrule
GPT-5.4 mini (low) & 97.1 & 8.8 & 27.1 & 25.0 \\
GPT-5.4 (low) & 100.0 & 73.5 & 68.8 & 68.8 \\
Sonnet 4.6 & 100.0 & 88.2 & 89.6 & 89.6 \\
Sonnet + thinking & 100.0 & 97.1 & 93.8 & 93.8 \\
\bottomrule
\end{tabular}
\caption{Paired raw controls (\%). K-chg/K-both: changed answers/both correct on 17 bases with changed truth. D-same/D-both: equal valid answers/both correct on 24 display-permuted bases. Rates average repetitions within base. Change alone does not establish correct sensitivity.}
\label{tab:controls}
\end{table}

%% file: diagnostic_findings.tex
Primary synthetic: among pooled raw and canonical requests, 70/139 valid outputs with both wrong starting stock and wrong future quantities exactly match replay from the emitted stock. There are 60 correct-stock/wrong-future outputs and 0 wrong-stock/correct-future outputs among 1536 valid requests. Historical loss is wrong despite correct starting stock in 71 requests. Across all arms, 0/2304 contracts are invalid and 0 responses hit the output-token limit (0 overlap).

Later knowledge: among pooled raw and canonical requests, 18/32 valid outputs with both wrong starting stock and wrong future quantities exactly match replay from the emitted stock. There are 15 correct-stock/wrong-future outputs and 0 wrong-stock/correct-future outputs among 384 valid requests. Historical loss is wrong despite correct starting stock in 17 requests. Across all arms, 0/576 contracts are invalid and 0 responses hit the output-token limit (0 overlap).

Display permutation: among raw requests, 16/35 valid outputs with both wrong starting stock and wrong future quantities exactly match replay from the emitted stock. There are 0 correct-stock/wrong-future outputs and 0 wrong-stock/correct-future outputs among 192 valid requests. Historical loss is wrong despite correct starting stock in 8 requests. Across all arms, 0/192 contracts are invalid and 0 responses hit the output-token limit (0 overlap).

Retail profiles: among pooled raw and canonical requests, 57/102 valid outputs with both wrong starting stock and wrong future quantities exactly match replay from the emitted stock. There are 56 correct-stock/wrong-future outputs and 1 wrong-stock/correct-future outputs among 767 valid requests. Historical loss is wrong despite correct starting stock in 35 requests. Across all arms, 1/1152 contracts are invalid and 0 responses hit the output-token limit (0 overlap).

These are descriptive counts over repeated requests, not independent failure probabilities. Replay agreement establishes observable consistency, not that the model internally used its emitted stock; correct future outputs after a wrong stock do not prove intentional recovery. Invalid and truncated counts overlap and must not be added.

%% file: continuation_full_table.tex
\begin{table}[htbp]
\centering\small
\begin{tabular}{llrr}
\toprule
Configuration & Input & Synthetic & Retail profiles \\
\midrule
GPT-5.4 mini (low) & raw & 37.5 / 50.5 & 12.5 / 14.6 \\
GPT-5.4 mini (low) & canonical & 76.0 / 88.0 & 46.9 / 90.6 \\
GPT-5.4 mini (low) & snapshot & 89.6 / 100.0 & 89.6 / 100.0 \\
GPT-5.4 (low) & raw & 88.0 / 91.1 & 90.6 / 90.6 \\
GPT-5.4 (low) & canonical & 98.4 / 99.5 & 91.7 / 100.0 \\
GPT-5.4 (low) & snapshot & 99.0 / 100.0 & 99.0 / 100.0 \\
Sonnet 4.6 & raw & 99.0 / 99.5 & 95.8 / 96.9 \\
Sonnet 4.6 & canonical & 99.5 / 100.0 & 99.0 / 100.0 \\
Sonnet 4.6 & snapshot & 100.0 / 100.0 & 100.0 / 100.0 \\
Sonnet + thinking & raw & 97.9 / 99.0 & 97.9 / 99.0 \\
Sonnet + thinking & canonical & 100.0 / 100.0 & 100.0 / 100.0 \\
Sonnet + thinking & snapshot & 100.0 / 100.0 & 100.0 / 100.0 \\
\bottomrule
\end{tabular}
\caption{Post-hoc future-joint accuracy (\%): original model rollout / executable continuation from its emitted stock. Synthetic and profile cells contain 192 and 96 original calls, respectively. Invalid five-field answers fail both endpoints. No state-only prompts were run; full-code replay is exact on every task.}
\label{tab:continuationfull}
\end{table}

%% file: extension_appendix.tex
\paragraph{Same-period low-effort replication.} The 576-call replication of mini's low-effort synthetic cells gives original versus replication future-joint accuracy of raw 37.5 vs 35.9\% (-1.6 [-9.4, +5.7] points), canonical 76.0 vs 67.2\% (-8.9 [-16.7, -1.0] points), snapshot 89.6 vs 89.1\% (-0.5 [-6.2, +5.2] points), and starting-stock accuracy of raw 50.5 vs 49.0\%, canonical 88.0 vs 84.4\%. The canonical cell replicates lower than the original, so collection-period variation of that size should be assumed when reading any single mini cell. Paired changes of the raised-effort cells relative to the replication baseline (future-joint, percentage points): GPT-5.4 mini (medium)/raw -9.4 [-17.2, -1.0]; GPT-5.4 mini (medium)/canonical -20.3 [-28.6, -12.0]; GPT-5.4 mini (medium)/snapshot +8.9 [+3.6, +14.6]; GPT-5.4 mini (high)/raw -10.9 [-19.8, -1.6]; GPT-5.4 mini (high)/canonical -32.3 [-42.2, -22.4]; GPT-5.4 mini (high)/snapshot +8.3 [+3.1, +14.1]. Paired changes of the state-only arms relative to the replication baseline: raw stock exact +6.2 [-1.0, +13.5]; raw future joint +19.3 [+10.9, +27.6]; canonical stock exact +9.4 [+3.6, +15.6]; canonical future joint +26.6 [+19.3, +34.4].

\paragraph{Character of the raised-effort canonical errors.} Among mini's wrong canonical starting stocks at high effort, 79 of 83 synthetic and 49 of 62 profile answers are underestimates, with median absolute errors of 32 and 14 units; 57 and 41 are exactly zero although the true stock is positive, 57 and 41 equal the stock obtained by replaying only the demands, and 13 and 13 equal the clipped net total. In twelve re-issued requests with reasoning summaries (diagnostic D1, supplement), the summaries lay out the correct selection and replay procedure, yet several answers are still zero or far too low, and one summary states that it must call the answer tool immediately and considers submitting placeholder zeros. This suggests that, under forced tool choice at raised effort, the model sometimes emits an answer before completing the replay it planned; the traces are illustrative, not a measured mechanism.

\paragraph{Code-inside arm for mini on synthetic histories, by history length.} tool-raw: 12 events 52 of 64 without code (43 exact), 0 wrong with code, 36 events 1 of 64 without code (0 exact), 2 wrong with code, 72 events 0 of 64 without code (0 exact), 4 wrong with code; tool-canonical: 12 events 51 of 64 without code (47 exact), 0 wrong with code, 36 events 2 of 64 without code (2 exact), 0 wrong with code, 72 events 0 of 64 without code (0 exact), 0 wrong with code. Calls answered without code concentrate on the shortest histories; the answers that were wrong despite running code are counted per stratum and were not individually inspected. The Anthropic configurations were not run with a code tool because their tool interface differs and both are at ceiling without it.

\begin{table}[htbp]
\centering\scriptsize
\setlength{\tabcolsep}{2pt}
\renewcommand{\arraystretch}{0.92}
\resizebox{\textwidth}{!}{%
\begin{tabular}{lllrr}
\toprule
Comparison (right minus left) & Metric & Bases & Difference & 95\% interval \\
\midrule
GPT-5.4 mini (low)/raw $\to$ canonical & future joint & 96 & +38.5 & [+28.6, +47.9] \\
GPT-5.4 mini (low)/raw $\to$ canonical & stock exact & 96 & +37.5 & [+28.6, +46.4] \\
GPT-5.4 mini (low)/raw $\to$ snapshot & future joint & 96 & +52.1 & [+43.8, +60.4] \\
GPT-5.4 mini (low)/raw $\to$ snapshot & stock exact & 96 & +49.5 & [+41.1, +58.3] \\
GPT-5.4 (low)/raw $\to$ canonical & future joint & 96 & +10.4 & [+5.7, +15.6] \\
GPT-5.4 (low)/raw $\to$ canonical & stock exact & 96 & +8.3 & [+3.6, +13.5] \\
GPT-5.4 (low)/raw $\to$ snapshot & future joint & 96 & +10.9 & [+6.2, +16.7] \\
GPT-5.4 (low)/raw $\to$ snapshot & stock exact & 96 & +8.9 & [+4.7, +14.1] \\
Sonnet 4.6/raw $\to$ canonical & future joint & 96 & +0.5 & [-1.0, +2.6] \\
Sonnet 4.6/raw $\to$ canonical & stock exact & 96 & +0.5 & [+0.0, +1.6] \\
Sonnet 4.6/raw $\to$ snapshot & future joint & 96 & +1.0 & [+0.0, +2.6] \\
Sonnet 4.6/raw $\to$ snapshot & stock exact & 96 & +0.5 & [+0.0, +1.6] \\
Sonnet + thinking/raw $\to$ canonical & future joint & 96 & +2.1 & [+0.5, +4.2] \\
Sonnet + thinking/raw $\to$ canonical & stock exact & 96 & +1.0 & [+0.0, +2.6] \\
Sonnet + thinking/raw $\to$ snapshot & future joint & 96 & +2.1 & [+0.5, +4.2] \\
Sonnet + thinking/raw $\to$ snapshot & stock exact & 96 & +1.0 & [+0.0, +2.6] \\
GPT-5.4 mini (low)/raw $\to$ GPT-5.4 mini (medium)/raw & future joint & 96 & -10.9 & [-18.8, -3.1] \\
GPT-5.4 mini (low)/raw $\to$ GPT-5.4 mini (medium)/raw & stock exact & 96 & -4.7 & [-13.0, +3.6] \\
GPT-5.4 mini (low)/canonical $\to$ GPT-5.4 mini (medium)/canonical & future joint & 96 & -29.2 & [-38.0, -20.3] \\
GPT-5.4 mini (low)/canonical $\to$ GPT-5.4 mini (medium)/canonical & stock exact & 96 & -21.9 & [-29.2, -14.6] \\
GPT-5.4 mini (low)/snapshot $\to$ GPT-5.4 mini (medium)/snapshot & future joint & 96 & +8.3 & [+4.2, +12.5] \\
GPT-5.4 mini (low)/snapshot $\to$ GPT-5.4 mini (medium)/snapshot & stock exact & 96 & +0.0 & [+0.0, +0.0] \\
GPT-5.4 mini (low)/raw $\to$ GPT-5.4 mini (high)/raw & future joint & 96 & -12.5 & [-21.9, -3.1] \\
GPT-5.4 mini (low)/raw $\to$ GPT-5.4 mini (high)/raw & stock exact & 96 & -8.9 & [-17.7, +0.5] \\
GPT-5.4 mini (low)/canonical $\to$ GPT-5.4 mini (high)/canonical & future joint & 96 & -41.1 & [-51.0, -31.2] \\
GPT-5.4 mini (low)/canonical $\to$ GPT-5.4 mini (high)/canonical & stock exact & 96 & -31.8 & [-40.6, -22.9] \\
GPT-5.4 mini (low)/snapshot $\to$ GPT-5.4 mini (high)/snapshot & future joint & 96 & +7.8 & [+3.1, +12.5] \\
GPT-5.4 mini (low)/snapshot $\to$ GPT-5.4 mini (high)/snapshot & stock exact & 96 & +0.0 & [+0.0, +0.0] \\
GPT-5.4 (low)/raw $\to$ GPT-5.4 (medium)/raw & future joint & 96 & +7.3 & [+1.0, +14.1] \\
GPT-5.4 (low)/raw $\to$ GPT-5.4 (medium)/raw & stock exact & 96 & +7.3 & [+2.6, +12.5] \\
GPT-5.4 (low)/canonical $\to$ GPT-5.4 (medium)/canonical & future joint & 96 & -2.6 & [-5.7, +0.5] \\
GPT-5.4 (low)/canonical $\to$ GPT-5.4 (medium)/canonical & stock exact & 96 & -1.6 & [-4.2, +0.5] \\
GPT-5.4 (low)/raw $\to$ GPT-5.4 (high)/raw & future joint & 96 & +6.2 & [+0.0, +13.0] \\
GPT-5.4 (low)/raw $\to$ GPT-5.4 (high)/raw & stock exact & 96 & +5.7 & [+0.5, +11.5] \\
GPT-5.4 (low)/canonical $\to$ GPT-5.4 (high)/canonical & future joint & 96 & -3.6 & [-7.8, +0.0] \\
GPT-5.4 (low)/canonical $\to$ GPT-5.4 (high)/canonical & stock exact & 96 & -1.0 & [-3.1, +1.0] \\
GPT-5.4 mini (low, auto tool choice)/raw $\to$ GPT-5.4 mini (high, auto tool choice)/raw & future joint & 96 & +22.4 & [+14.1, +30.7] \\
GPT-5.4 mini (low)/raw $\to$ GPT-5.4 mini (low, auto tool choice)/raw & future joint & 96 & +19.3 & [+11.5, +27.1] \\
GPT-5.4 mini (high)/raw $\to$ GPT-5.4 mini (high, auto tool choice)/raw & future joint & 96 & +54.2 & [+44.3, +63.0] \\
GPT-5.4 mini (low, auto tool choice)/raw $\to$ GPT-5.4 mini (high, auto tool choice)/raw & stock exact & 96 & +23.4 & [+15.1, +31.2] \\
GPT-5.4 mini (low)/raw $\to$ GPT-5.4 mini (low, auto tool choice)/raw & stock exact & 96 & +8.3 & [+2.1, +15.1] \\
GPT-5.4 mini (high)/raw $\to$ GPT-5.4 mini (high, auto tool choice)/raw & stock exact & 96 & +40.6 & [+31.2, +49.5] \\
GPT-5.4 mini (low, auto tool choice)/canonical $\to$ GPT-5.4 mini (high, auto tool choice)/canonical & future joint & 96 & +6.8 & [+0.5, +13.5] \\
GPT-5.4 mini (low)/canonical $\to$ GPT-5.4 mini (low, auto tool choice)/canonical & future joint & 96 & +10.4 & [+3.1, +18.2] \\
GPT-5.4 mini (high)/canonical $\to$ GPT-5.4 mini (high, auto tool choice)/canonical & future joint & 96 & +58.3 & [+50.0, +66.7] \\
GPT-5.4 mini (low, auto tool choice)/canonical $\to$ GPT-5.4 mini (high, auto tool choice)/canonical & stock exact & 96 & +5.2 & [+0.5, +10.4] \\
GPT-5.4 mini (low)/canonical $\to$ GPT-5.4 mini (low, auto tool choice)/canonical & stock exact & 96 & +3.6 & [-1.6, +8.9] \\
GPT-5.4 mini (high)/canonical $\to$ GPT-5.4 mini (high, auto tool choice)/canonical & stock exact & 96 & +40.6 & [+31.8, +49.5] \\
GPT-5.4 mini (low)/raw $\to$ state-raw & stock exact & 96 & +4.7 & [-2.1, +11.5] \\
GPT-5.4 mini (low)/raw $\to$ state-raw & future joint & 96 & +17.7 & [+9.9, +25.5] \\
GPT-5.4 mini (low)/canonical $\to$ state-canonical & stock exact & 96 & +5.7 & [+0.0, +11.5] \\
GPT-5.4 mini (low)/canonical $\to$ state-canonical & future joint & 96 & +17.7 & [+9.9, +25.5] \\
GPT-5.4 mini (high)/raw $\to$ state-raw & stock exact & 96 & +12.5 & [+3.6, +20.8] \\
GPT-5.4 mini (high)/raw $\to$ state-raw & future joint & 96 & +29.2 & [+19.8, +38.5] \\
GPT-5.4 mini (high)/canonical $\to$ state-canonical & stock exact & 96 & +15.1 & [+7.3, +22.9] \\
GPT-5.4 mini (high)/canonical $\to$ state-canonical & future joint & 96 & +36.5 & [+26.6, +46.4] \\
GPT-5.4 (low)/raw $\to$ state-raw & stock exact & 96 & +0.5 & [-5.2, +6.2] \\
GPT-5.4 (low)/raw $\to$ state-raw & future joint & 96 & +3.6 & [-2.6, +9.9] \\
GPT-5.4 (low)/canonical $\to$ state-canonical & stock exact & 96 & -1.0 & [-3.1, +1.0] \\
GPT-5.4 (low)/canonical $\to$ state-canonical & future joint & 96 & +0.0 & [-2.6, +2.6] \\
Sonnet 4.6/raw $\to$ state-raw & stock exact & 96 & +0.5 & [+0.0, +1.6] \\
Sonnet 4.6/raw $\to$ state-raw & future joint & 96 & +1.0 & [+0.0, +2.6] \\
Sonnet 4.6/canonical $\to$ state-canonical & stock exact & 96 & +0.0 & [+0.0, +0.0] \\
Sonnet 4.6/canonical $\to$ state-canonical & future joint & 96 & +0.5 & [+0.0, +1.6] \\
Sonnet + thinking/raw $\to$ state-raw & stock exact & 96 & +1.0 & [+0.0, +2.6] \\
Sonnet + thinking/raw $\to$ state-raw & future joint & 96 & +2.1 & [+0.5, +4.2] \\
Sonnet + thinking/canonical $\to$ state-canonical & stock exact & 96 & +0.0 & [+0.0, +0.0] \\
Sonnet + thinking/canonical $\to$ state-canonical & future joint & 96 & +0.0 & [+0.0, +0.0] \\
GPT-5.4 mini (low)/raw $\to$ tool-raw & future joint & 96 & +54.2 & [+45.3, +62.5] \\
GPT-5.4 mini (low)/raw $\to$ tool-raw & stock exact & 96 & +44.3 & [+35.4, +53.1] \\
GPT-5.4 mini (low)/state-raw $\to$ tool-raw & future joint & 96 & +36.5 & [+27.1, +45.8] \\
GPT-5.4 mini (low)/canonical $\to$ tool-canonical & future joint & 96 & +21.9 & [+14.6, +29.2] \\
GPT-5.4 mini (low)/canonical $\to$ tool-canonical & stock exact & 96 & +11.5 & [+6.2, +16.7] \\
GPT-5.4 mini (low)/state-canonical $\to$ tool-canonical & future joint & 96 & +4.2 & [+0.5, +8.3] \\
GPT-5.4 (low)/raw $\to$ tool-raw & future joint & 96 & +12.0 & [+7.3, +17.2] \\
GPT-5.4 (low)/raw $\to$ tool-raw & stock exact & 96 & +8.9 & [+4.7, +14.1] \\
GPT-5.4 (low)/state-raw $\to$ tool-raw & future joint & 96 & +8.3 & [+4.7, +12.5] \\
GPT-5.4 (low)/canonical $\to$ tool-canonical & future joint & 96 & -1.0 & [-4.2, +1.6] \\
GPT-5.4 (low)/canonical $\to$ tool-canonical & stock exact & 96 & -1.0 & [-3.1, +1.0] \\
GPT-5.4 (low)/state-canonical $\to$ tool-canonical & future joint & 96 & -1.0 & [-4.2, +1.6] \\
GPT-5.4 mini (low, auto tool choice)/raw $\to$ GPT-5.4 mini (low)/tool-raw & future joint & 96 & +34.9 & [+26.0, +43.8] \\
GPT-5.4 mini (low, auto tool choice)/raw $\to$ GPT-5.4 mini (low)/tool-raw & stock exact & 96 & +35.9 & [+27.1, +44.8] \\
GPT-5.4 mini (low, auto tool choice)/canonical $\to$ GPT-5.4 mini (low)/tool-canonical & future joint & 96 & +11.5 & [+6.2, +17.7] \\
GPT-5.4 mini (low, auto tool choice)/canonical $\to$ GPT-5.4 mini (low)/tool-canonical & stock exact & 96 & +7.8 & [+3.6, +12.5] \\
\bottomrule
\end{tabular}}
\caption{Paired differences in percentage points, synthetic cohort: matched by history and repetition index, averaged within history; descriptive 95\% history-bootstrap percentile intervals (10,000 draws), not multiplicity adjusted. The same-period replication moved one identical cell by $-8.9$ points, so differences of that size are within run-to-run variation. Cells at or near 100\% give degenerate intervals. Identical rows arise when two comparisons have identical per-history differences.}
\label{tab:extpaired_heldout}
\end{table}

\begin{table}[htbp]
\centering\scriptsize
\setlength{\tabcolsep}{2pt}
\renewcommand{\arraystretch}{0.92}
\resizebox{\textwidth}{!}{%
\begin{tabular}{lllrr}
\toprule
Comparison (right minus left) & Metric & Bases & Difference & 95\% interval \\
\midrule
GPT-5.4 mini (low)/raw $\to$ canonical & future joint & 48 & +34.4 & [+24.0, +44.8] \\
GPT-5.4 mini (low)/raw $\to$ canonical & stock exact & 48 & +76.0 & [+66.7, +84.4] \\
GPT-5.4 mini (low)/raw $\to$ snapshot & future joint & 48 & +77.1 & [+68.8, +85.4] \\
GPT-5.4 mini (low)/raw $\to$ snapshot & stock exact & 48 & +85.4 & [+78.1, +91.7] \\
GPT-5.4 (low)/raw $\to$ canonical & future joint & 48 & +1.0 & [-6.2, +8.3] \\
GPT-5.4 (low)/raw $\to$ canonical & stock exact & 48 & +9.4 & [+4.2, +15.6] \\
GPT-5.4 (low)/raw $\to$ snapshot & future joint & 48 & +8.3 & [+3.1, +14.6] \\
GPT-5.4 (low)/raw $\to$ snapshot & stock exact & 48 & +9.4 & [+4.2, +15.6] \\
Sonnet 4.6/raw $\to$ canonical & future joint & 48 & +3.1 & [+0.0, +8.3] \\
Sonnet 4.6/raw $\to$ canonical & stock exact & 48 & +3.1 & [+0.0, +8.3] \\
Sonnet 4.6/raw $\to$ snapshot & future joint & 48 & +4.2 & [+0.0, +9.4] \\
Sonnet 4.6/raw $\to$ snapshot & stock exact & 48 & +3.1 & [+0.0, +8.3] \\
Sonnet + thinking/raw $\to$ canonical & future joint & 48 & +2.1 & [+0.0, +6.2] \\
Sonnet + thinking/raw $\to$ canonical & stock exact & 48 & +1.0 & [+0.0, +3.1] \\
Sonnet + thinking/raw $\to$ snapshot & future joint & 48 & +2.1 & [+0.0, +6.2] \\
Sonnet + thinking/raw $\to$ snapshot & stock exact & 48 & +1.0 & [+0.0, +3.1] \\
GPT-5.4 mini (low)/raw $\to$ GPT-5.4 mini (medium)/raw & future joint & 48 & +4.2 & [-4.2, +12.5] \\
GPT-5.4 mini (low)/raw $\to$ GPT-5.4 mini (medium)/raw & stock exact & 48 & +9.4 & [+1.0, +18.8] \\
GPT-5.4 mini (low)/canonical $\to$ GPT-5.4 mini (medium)/canonical & future joint & 48 & -8.3 & [-20.8, +4.2] \\
GPT-5.4 mini (low)/canonical $\to$ GPT-5.4 mini (medium)/canonical & stock exact & 48 & -38.5 & [-50.0, -27.1] \\
GPT-5.4 mini (low)/snapshot $\to$ GPT-5.4 mini (medium)/snapshot & future joint & 48 & +5.2 & [-1.0, +11.5] \\
GPT-5.4 mini (low)/snapshot $\to$ GPT-5.4 mini (medium)/snapshot & stock exact & 48 & +0.0 & [+0.0, +0.0] \\
GPT-5.4 mini (low)/raw $\to$ GPT-5.4 mini (high)/raw & future joint & 48 & +8.3 & [-1.0, +17.7] \\
GPT-5.4 mini (low)/raw $\to$ GPT-5.4 mini (high)/raw & stock exact & 48 & +15.6 & [+6.2, +25.0] \\
GPT-5.4 mini (low)/canonical $\to$ GPT-5.4 mini (high)/canonical & future joint & 48 & -17.7 & [-30.2, -5.2] \\
GPT-5.4 mini (low)/canonical $\to$ GPT-5.4 mini (high)/canonical & stock exact & 48 & -55.2 & [-65.6, -44.8] \\
GPT-5.4 mini (low)/snapshot $\to$ GPT-5.4 mini (high)/snapshot & future joint & 48 & +4.2 & [-4.2, +12.5] \\
GPT-5.4 mini (low)/snapshot $\to$ GPT-5.4 mini (high)/snapshot & stock exact & 48 & +0.0 & [+0.0, +0.0] \\
GPT-5.4 (low)/raw $\to$ GPT-5.4 (medium)/raw & future joint & 48 & +5.2 & [-2.1, +12.5] \\
GPT-5.4 (low)/raw $\to$ GPT-5.4 (medium)/raw & stock exact & 48 & +5.2 & [-2.1, +12.5] \\
GPT-5.4 (low)/canonical $\to$ GPT-5.4 (medium)/canonical & future joint & 48 & +3.1 & [-4.2, +10.4] \\
GPT-5.4 (low)/canonical $\to$ GPT-5.4 (medium)/canonical & stock exact & 48 & -1.0 & [-3.1, +0.0] \\
GPT-5.4 (low)/raw $\to$ GPT-5.4 (high)/raw & future joint & 48 & +8.3 & [+2.1, +14.6] \\
GPT-5.4 (low)/raw $\to$ GPT-5.4 (high)/raw & stock exact & 48 & +8.3 & [+2.1, +14.6] \\
GPT-5.4 (low)/canonical $\to$ GPT-5.4 (high)/canonical & future joint & 48 & -3.1 & [-11.5, +4.2] \\
GPT-5.4 (low)/canonical $\to$ GPT-5.4 (high)/canonical & stock exact & 48 & -9.4 & [-15.6, -4.2] \\
GPT-5.4 mini (low, auto tool choice)/raw $\to$ GPT-5.4 mini (high, auto tool choice)/raw & future joint & 48 & +34.4 & [+21.9, +46.9] \\
GPT-5.4 mini (low)/raw $\to$ GPT-5.4 mini (low, auto tool choice)/raw & future joint & 48 & +2.1 & [-6.2, +10.4] \\
GPT-5.4 mini (high)/raw $\to$ GPT-5.4 mini (high, auto tool choice)/raw & future joint & 48 & +28.1 & [+16.7, +39.6] \\
GPT-5.4 mini (low, auto tool choice)/raw $\to$ GPT-5.4 mini (high, auto tool choice)/raw & stock exact & 48 & +30.2 & [+18.8, +41.7] \\
GPT-5.4 mini (low)/raw $\to$ GPT-5.4 mini (low, auto tool choice)/raw & stock exact & 48 & +4.2 & [-5.2, +13.5] \\
GPT-5.4 mini (high)/raw $\to$ GPT-5.4 mini (high, auto tool choice)/raw & stock exact & 48 & +18.8 & [+7.3, +30.2] \\
GPT-5.4 mini (low, auto tool choice)/canonical $\to$ GPT-5.4 mini (high, auto tool choice)/canonical & future joint & 48 & +12.5 & [+4.2, +19.8] \\
GPT-5.4 mini (low)/canonical $\to$ GPT-5.4 mini (low, auto tool choice)/canonical & future joint & 48 & +35.4 & [+24.0, +46.9] \\
GPT-5.4 mini (high)/canonical $\to$ GPT-5.4 mini (high, auto tool choice)/canonical & future joint & 48 & +65.6 & [+56.2, +75.0] \\
GPT-5.4 mini (low, auto tool choice)/canonical $\to$ GPT-5.4 mini (high, auto tool choice)/canonical & stock exact & 48 & +4.2 & [-2.1, +10.4] \\
GPT-5.4 mini (low)/canonical $\to$ GPT-5.4 mini (low, auto tool choice)/canonical & stock exact & 48 & +2.1 & [-6.2, +11.5] \\
GPT-5.4 mini (high)/canonical $\to$ GPT-5.4 mini (high, auto tool choice)/canonical & stock exact & 48 & +61.5 & [+51.0, +71.9] \\
GPT-5.4 mini (low)/raw $\to$ state-raw & stock exact & 48 & +3.1 & [-5.2, +11.5] \\
GPT-5.4 mini (low)/raw $\to$ state-raw & future joint & 48 & +5.2 & [-3.1, +13.5] \\
GPT-5.4 mini (low)/canonical $\to$ state-canonical & stock exact & 48 & +8.3 & [+3.1, +14.6] \\
GPT-5.4 mini (low)/canonical $\to$ state-canonical & future joint & 48 & +52.1 & [+42.7, +61.5] \\
GPT-5.4 mini (high)/raw $\to$ state-raw & stock exact & 48 & +4.2 & [-8.3, +16.7] \\
GPT-5.4 mini (high)/raw $\to$ state-raw & future joint & 48 & +13.5 & [+0.0, +27.1] \\
GPT-5.4 mini (high)/canonical $\to$ state-canonical & stock exact & 48 & -1.0 & [-11.5, +9.4] \\
GPT-5.4 mini (high)/canonical $\to$ state-canonical & future joint & 48 & +5.2 & [-7.3, +17.7] \\
GPT-5.4 (low)/raw $\to$ state-raw & stock exact & 48 & +0.0 & [-8.3, +8.3] \\
GPT-5.4 (low)/raw $\to$ state-raw & future joint & 48 & +0.0 & [-8.3, +8.3] \\
GPT-5.4 (low)/canonical $\to$ state-canonical & stock exact & 48 & -1.0 & [-3.1, +0.0] \\
GPT-5.4 (low)/canonical $\to$ state-canonical & future joint & 48 & +7.3 & [+2.1, +13.5] \\
Sonnet 4.6/raw $\to$ state-raw & stock exact & 48 & -1.0 & [-3.1, +0.0] \\
Sonnet 4.6/raw $\to$ state-raw & future joint & 48 & +0.0 & [-3.1, +3.1] \\
Sonnet 4.6/canonical $\to$ state-canonical & stock exact & 48 & +0.0 & [+0.0, +0.0] \\
Sonnet 4.6/canonical $\to$ state-canonical & future joint & 48 & +1.0 & [+0.0, +3.1] \\
Sonnet + thinking/raw $\to$ state-raw & stock exact & 48 & +1.0 & [+0.0, +3.1] \\
Sonnet + thinking/raw $\to$ state-raw & future joint & 48 & +2.1 & [+0.0, +6.2] \\
Sonnet + thinking/canonical $\to$ state-canonical & stock exact & 48 & +0.0 & [+0.0, +0.0] \\
Sonnet + thinking/canonical $\to$ state-canonical & future joint & 48 & +0.0 & [+0.0, +0.0] \\
GPT-5.4 mini (low)/raw $\to$ tool-raw & future joint & 48 & +77.1 & [+68.8, +84.4] \\
GPT-5.4 mini (low)/raw $\to$ tool-raw & stock exact & 48 & +75.0 & [+66.7, +83.3] \\
GPT-5.4 mini (low)/state-raw $\to$ tool-raw & future joint & 48 & +71.9 & [+60.4, +82.3] \\
GPT-5.4 mini (low)/canonical $\to$ tool-canonical & future joint & 48 & +53.1 & [+43.8, +61.5] \\
GPT-5.4 mini (low)/canonical $\to$ tool-canonical & stock exact & 48 & +9.4 & [+4.2, +16.7] \\
GPT-5.4 mini (low)/state-canonical $\to$ tool-canonical & future joint & 48 & +1.0 & [+0.0, +3.1] \\
GPT-5.4 (low)/raw $\to$ tool-raw & future joint & 48 & +9.4 & [+4.2, +15.6] \\
GPT-5.4 (low)/raw $\to$ tool-raw & stock exact & 48 & +9.4 & [+4.2, +15.6] \\
GPT-5.4 (low)/state-raw $\to$ tool-raw & future joint & 48 & +9.4 & [+4.2, +15.6] \\
GPT-5.4 (low)/canonical $\to$ tool-canonical & future joint & 48 & +8.3 & [+3.1, +13.5] \\
GPT-5.4 (low)/canonical $\to$ tool-canonical & stock exact & 48 & +0.0 & [+0.0, +0.0] \\
GPT-5.4 (low)/state-canonical $\to$ tool-canonical & future joint & 48 & +1.0 & [+0.0, +3.1] \\
GPT-5.4 mini (low, auto tool choice)/raw $\to$ GPT-5.4 mini (low)/tool-raw & future joint & 48 & +75.0 & [+64.6, +84.4] \\
GPT-5.4 mini (low, auto tool choice)/raw $\to$ GPT-5.4 mini (low)/tool-raw & stock exact & 48 & +70.8 & [+60.4, +81.2] \\
GPT-5.4 mini (low, auto tool choice)/canonical $\to$ GPT-5.4 mini (low)/tool-canonical & future joint & 48 & +17.7 & [+10.4, +25.0] \\
GPT-5.4 mini (low, auto tool choice)/canonical $\to$ GPT-5.4 mini (low)/tool-canonical & stock exact & 48 & +7.3 & [+3.1, +12.5] \\
\bottomrule
\end{tabular}}
\caption{Paired differences in percentage points, retail profiles cohort: matched by history and repetition index, averaged within history; descriptive 95\% history-bootstrap percentile intervals (10,000 draws), not multiplicity adjusted. The same-period replication moved one identical cell by $-8.9$ points, so differences of that size are within run-to-run variation. Cells at or near 100\% give degenerate intervals. Identical rows arise when two comparisons have identical per-history differences.}
\label{tab:extpaired_profiles}
\end{table}

%% file: ladder_appendix.tex
\paragraph{Transport.} Of the 23,808 Amendment~4 calls, 637 needed a bounded transport retry (HTTP 429/5xx before any completion existed; attempt counts are stored per record and no model output was re-sampled), and 0 open-weight calls were served by a provider other than the pinned one. Requested effort for the open-weight models is OpenRouter's unified \texttt{reasoning.effort}, which each upstream maps onto its own control.

\paragraph{Same-period GPT-5.4 forced-call replication (E6b).} Future-joint accuracy on synthetic histories, snapshot: original 99.0, same-period replication 97.9, automatic choice 91.1\%; state-canonical: original 98.4, same-period replication 99.0, automatic choice 83.9\%. Paired drift, replication minus original forced cell: snapshot -1.0 [-3.6, +1.6]; state-canonical +0.5 [-1.6, +2.6]. Interface effect against the contemporaneous baseline, automatic minus replicated forced call: snapshot -6.8 [-11.5, -2.6]; state-canonical -15.1 [-21.4, -9.4].

\begin{table}[htbp]
\centering\scriptsize
\setlength{\tabcolsep}{3pt}
\begin{tabular}{llrrrrrr}
\toprule
Model & Configuration & Arm & Future joint & Missing & Truncated & Mean reasoning tok. & Retried calls \\
\midrule
GPT-5.4 mini & low, forced & raw & 37.5 & 0 & 0 & 790 &  \\
 & low, forced & canonical & 76.0 & 0 & 0 & 515 &  \\
 & high, forced & raw & 25.0 & 0 & 0 & 1214 &  \\
 & high, forced & canonical & 34.9 & 0 & 0 & 708 &  \\
 & low, auto & raw & 56.8 & 1 & 0 & 975 &  \\
 & low, auto & canonical & 86.5 & 0 & 0 & 565 &  \\
 & high, auto & raw & 79.2 & 1 & 1 & 3054 &  \\
 & high, auto & canonical & 93.2 & 0 & 0 & 1425 &  \\
\midrule
GPT-5.4 & low, forced & raw & 88.0 & 0 & 0 & 710 &  \\
 & low, forced & canonical & 98.4 & 0 & 0 & 419 &  \\
 & high, forced & raw & 94.3 & 0 & 0 & 1602 &  \\
 & high, forced & canonical & 94.8 & 0 & 0 & 749 &  \\
 & low, auto & raw & 98.4 & 0 & 0 & 707 &  \\
 & low, auto & canonical & 98.4 & 0 & 0 & 415 &  \\
 & high, auto & raw & 99.0 & 0 & 0 & 1554 &  \\
 & high, auto & canonical & 100.0 & 0 & 0 & 688 &  \\
\midrule
gpt-oss-20b & low, forced & raw & 23.4 & 85 & 0 & 494 &  \\
 & low, forced & canonical & 52.1 & 59 & 0 & 482 &  \\
 & high, forced & raw & 42.2 & 106 & 0 & 10545 &  \\
 & high, forced & canonical & 66.1 & 65 & 0 & 9862 &  \\
 & low, auto & raw & 24.0 & 88 & 0 & 498 &  \\
 & low, auto & canonical & 51.0 & 51 & 0 & 598 &  \\
 & high, auto & raw & 39.1 & 112 & 97 & 12678 &  \\
 & high, auto & canonical & 59.4 & 75 & 58 & 11037 &  \\
\midrule
gpt-oss-120b & low, forced & raw & 45.8 & 0 & 0 & 941 &  \\
 & low, forced & canonical & 69.3 & 0 & 0 & 664 &  \\
 & high, forced & raw & 53.6 & 89 & 89 & 6539 &  \\
 & high, forced & canonical & 66.1 & 65 & 65 & 5925 &  \\
 & low, auto & raw & 49.5 & 41 & 0 & 1023 &  \\
 & low, auto & canonical & 73.4 & 12 & 0 & 713 &  \\
 & high, auto & raw & 49.0 & 98 & 98 & 6679 &  \\
 & high, auto & canonical & 68.8 & 60 & 60 & 5986 &  \\
\midrule
Qwen3.8-27B & low, forced & raw & 76.6 & 40 & 0 & 2379 &  \\
 & low, forced & canonical & 91.7 & 16 & 0 & 1607 &  \\
 & high, forced & raw & 80.2 & 28 & 0 & 2103 &  \\
 & high, forced & canonical & 91.7 & 10 & 0 & 1225 &  \\
 & low, auto & raw & 95.3 & 3 & 1 & 2679 &  \\
 & low, auto & canonical & 99.5 & 0 & 0 & 1683 &  \\
 & high, auto & raw & 88.5 & 19 & 4 & 2271 &  \\
 & high, auto & canonical & 92.7 & 8 & 0 & 1212 &  \\
\midrule
DeepSeek V4 Flash & low, forced & raw & 2.1 & 4 & 0 & 103 &  \\
 & low, forced & canonical & 2.1 & 6 & 0 & 32 &  \\
 & high, forced & raw & 1.6 & 10 & 0 & 78 &  \\
 & high, forced & canonical & 1.0 & 15 & 0 & 24 &  \\
 & low, auto & raw & 98.4 & 0 & 0 & 3443 &  \\
 & low, auto & canonical & 99.0 & 0 & 0 & 1695 &  \\
 & high, auto & raw & 93.2 & 2 & 2 & 4756 &  \\
 & high, auto & canonical & 97.4 & 0 & 0 & 2451 &  \\
\midrule
GLM-5.3 Flash & low, forced & raw & 52.6 & 0 & 0 & 444 &  \\
 & low, forced & canonical & 62.5 & 0 & 0 & 316 &  \\
 & high, forced & raw & 74.5 & 0 & 0 & 823 &  \\
 & high, forced & canonical & 82.3 & 0 & 0 & 415 &  \\
 & low, auto & raw & 55.7 & 0 & 0 & 441 &  \\
 & low, auto & canonical & 64.1 & 2 & 1 & 441 &  \\
 & high, auto & raw & 76.6 & 1 & 0 & 991 &  \\
 & high, auto & canonical & 84.4 & 0 & 0 & 503 &  \\
\midrule
DeepSeek V4 Pro & low, forced & raw & 96.4 & 6 & 0 & 3662 &  \\
 & low, forced & canonical & 98.4 & 2 & 0 & 2523 &  \\
 & high, forced & raw & 91.7 & 16 & 9 & 5211 &  \\
 & high, forced & canonical & 82.3 & 32 & 1 & 3191 &  \\
 & low, auto & raw & 99.0 & 0 & 0 & 4142 &  \\
 & low, auto & canonical & 99.0 & 0 & 0 & 2461 &  \\
 & high, auto & raw & 93.2 & 12 & 11 & 5768 &  \\
 & high, auto & canonical & 97.4 & 2 & 2 & 3971 &  \\
\bottomrule
\end{tabular}
\caption{Effort $\times$ interface cells on the synthetic cohort (192 calls each): future-joint accuracy (\%), calls with no answer (including refusals), calls truncated at the 16,384-token output limit, and mean provider-reported reasoning tokens. Each upstream maps the unified effort parameter onto its own control, so effort levels are comparable within a model family only.}
\label{tab:ladder_cells}
\end{table}

\begin{table}[htbp]
\centering\scriptsize
\setlength{\tabcolsep}{2pt}
\resizebox{\textwidth}{!}{%
\begin{tabular}{lllrrrrrrr}
\toprule
Cohort & Configuration & Arm & refusal & text only & other tool & multiple calls & stream error & truncated & empty \\
\midrule
Synthetic & DeepSeek V4 Flash (high, forced call) & canonical & 0 & 0 & 0 & 11 & 0 & 4 & 0 \\
Synthetic & DeepSeek V4 Flash (high, forced call) & raw & 0 & 0 & 0 & 8 & 0 & 2 & 0 \\
Synthetic & DeepSeek V4 Flash (low, forced call) & canonical & 0 & 0 & 0 & 6 & 0 & 0 & 0 \\
Synthetic & DeepSeek V4 Flash (low, forced call) & raw & 0 & 0 & 0 & 4 & 0 & 0 & 0 \\
Synthetic & DeepSeek V4 Pro (high, forced call) & canonical & 0 & 0 & 0 & 0 & 0 & 1 & 31 \\
Synthetic & DeepSeek V4 Pro (high, forced call) & raw & 0 & 0 & 0 & 0 & 0 & 9 & 7 \\
Synthetic & DeepSeek V4 Pro (high, auto tool choice) & raw & 0 & 0 & 0 & 0 & 0 & 12 & 0 \\
Synthetic & DeepSeek V4 Pro (low, forced call) & raw & 0 & 0 & 0 & 0 & 0 & 0 & 6 \\
Synthetic & Haiku 4.5 + thinking & raw & 0 & 0 & 0 & 0 & 0 & 10 & 0 \\
Synthetic & Haiku 4.5 + thinking & state-raw & 0 & 0 & 0 & 0 & 0 & 11 & 0 \\
Synthetic & gpt-oss-120b (high, forced call) & canonical & 0 & 0 & 0 & 0 & 0 & 65 & 0 \\
Synthetic & gpt-oss-120b (high, forced call) & raw & 0 & 0 & 0 & 0 & 0 & 89 & 0 \\
Synthetic & gpt-oss-120b (high, auto tool choice) & canonical & 0 & 0 & 0 & 0 & 0 & 60 & 0 \\
Synthetic & gpt-oss-120b (high, auto tool choice) & raw & 0 & 0 & 0 & 0 & 0 & 98 & 0 \\
Synthetic & gpt-oss-120b (low, auto tool choice) & canonical & 7 & 5 & 0 & 0 & 0 & 0 & 0 \\
Synthetic & gpt-oss-120b (low, auto tool choice) & raw & 24 & 17 & 0 & 0 & 0 & 0 & 0 \\
Synthetic & gpt-oss-120b (low, auto tool choice) & state-canonical & 1 & 3 & 0 & 0 & 0 & 0 & 0 \\
Synthetic & gpt-oss-120b (low, auto tool choice) & state-raw & 9 & 7 & 0 & 0 & 0 & 0 & 0 \\
Synthetic & gpt-oss-20b (high, forced call) & canonical & 0 & 0 & 9 & 0 & 56 & 0 & 0 \\
Synthetic & gpt-oss-20b (high, forced call) & raw & 0 & 0 & 10 & 0 & 96 & 0 & 0 \\
Synthetic & gpt-oss-20b (high, auto tool choice) & canonical & 0 & 2 & 12 & 0 & 2 & 58 & 1 \\
Synthetic & gpt-oss-20b (high, auto tool choice) & raw & 0 & 1 & 11 & 0 & 0 & 97 & 3 \\
Synthetic & gpt-oss-20b (low, forced call) & canonical & 0 & 0 & 7 & 0 & 52 & 0 & 0 \\
Synthetic & gpt-oss-20b (low, forced call) & raw & 0 & 0 & 7 & 0 & 78 & 0 & 0 \\
Synthetic & gpt-oss-20b (low, auto tool choice) & canonical & 37 & 1 & 5 & 0 & 8 & 0 & 0 \\
Synthetic & gpt-oss-20b (low, auto tool choice) & raw & 70 & 1 & 6 & 0 & 11 & 0 & 0 \\
Synthetic & gpt-oss-20b (low, auto tool choice) & state-canonical & 18 & 1 & 12 & 0 & 1 & 0 & 0 \\
Synthetic & gpt-oss-20b (low, auto tool choice) & state-raw & 27 & 2 & 19 & 0 & 5 & 0 & 0 \\
Synthetic & Qwen3.8-27B (high, forced call) & canonical & 0 & 0 & 0 & 0 & 10 & 0 & 0 \\
Synthetic & Qwen3.8-27B (high, forced call) & raw & 0 & 0 & 0 & 0 & 28 & 0 & 0 \\
Synthetic & Qwen3.8-27B (high, auto tool choice) & canonical & 0 & 0 & 0 & 0 & 8 & 0 & 0 \\
Synthetic & Qwen3.8-27B (high, auto tool choice) & raw & 0 & 0 & 0 & 0 & 15 & 4 & 0 \\
Synthetic & Qwen3.8-27B (low, forced call) & canonical & 0 & 0 & 0 & 0 & 16 & 0 & 0 \\
Synthetic & Qwen3.8-27B (low, forced call) & raw & 0 & 0 & 0 & 0 & 40 & 0 & 0 \\
Synthetic & Qwen3.8-27B (low, auto tool choice) & raw & 0 & 0 & 0 & 0 & 1 & 1 & 1 \\
Synthetic & Qwen3.8-27B (low, auto tool choice) & state-raw & 0 & 0 & 0 & 0 & 0 & 3 & 0 \\
Retail profiles & DeepSeek V4 Flash (high, forced call) & raw & 0 & 0 & 0 & 6 & 0 & 2 & 0 \\
Retail profiles & DeepSeek V4 Flash (high, auto tool choice) & raw & 0 & 0 & 2 & 0 & 0 & 1 & 0 \\
Retail profiles & DeepSeek V4 Flash (low, forced call) & canonical & 0 & 0 & 0 & 3 & 0 & 1 & 0 \\
Retail profiles & DeepSeek V4 Pro (high, forced call) & canonical & 0 & 0 & 0 & 0 & 0 & 1 & 19 \\
Retail profiles & DeepSeek V4 Pro (high, forced call) & raw & 0 & 0 & 0 & 0 & 0 & 4 & 15 \\
Retail profiles & DeepSeek V4 Pro (high, auto tool choice) & raw & 0 & 0 & 0 & 0 & 0 & 3 & 0 \\
Retail profiles & DeepSeek V4 Pro (low, forced call) & raw & 0 & 0 & 0 & 0 & 0 & 0 & 7 \\
Retail profiles & Haiku 4.5 + thinking & raw & 0 & 0 & 0 & 0 & 0 & 37 & 0 \\
Retail profiles & Haiku 4.5 + thinking & state-raw & 0 & 0 & 0 & 0 & 0 & 21 & 0 \\
Retail profiles & gpt-oss-120b (high, forced call) & canonical & 0 & 0 & 1 & 0 & 0 & 84 & 0 \\
Retail profiles & gpt-oss-120b (high, forced call) & raw & 0 & 0 & 0 & 0 & 0 & 96 & 0 \\
Retail profiles & gpt-oss-120b (high, auto tool choice) & canonical & 0 & 0 & 0 & 0 & 0 & 86 & 0 \\
Retail profiles & gpt-oss-120b (high, auto tool choice) & raw & 0 & 0 & 0 & 0 & 0 & 96 & 0 \\
Retail profiles & gpt-oss-120b (low, auto tool choice) & raw & 17 & 30 & 0 & 0 & 0 & 0 & 0 \\
Retail profiles & gpt-oss-20b (high, forced call) & canonical & 0 & 0 & 1 & 0 & 48 & 0 & 0 \\
Retail profiles & gpt-oss-20b (high, forced call) & raw & 0 & 0 & 0 & 0 & 96 & 0 & 0 \\
Retail profiles & gpt-oss-20b (high, auto tool choice) & canonical & 0 & 0 & 3 & 0 & 1 & 46 & 0 \\
Retail profiles & gpt-oss-20b (high, auto tool choice) & raw & 0 & 0 & 0 & 0 & 0 & 96 & 0 \\
Retail profiles & gpt-oss-20b (low, forced call) & canonical & 0 & 0 & 9 & 0 & 44 & 0 & 0 \\
Retail profiles & gpt-oss-20b (low, forced call) & raw & 0 & 0 & 6 & 0 & 67 & 0 & 0 \\
Retail profiles & gpt-oss-20b (low, auto tool choice) & canonical & 44 & 2 & 7 & 0 & 5 & 0 & 0 \\
Retail profiles & gpt-oss-20b (low, auto tool choice) & raw & 55 & 2 & 8 & 0 & 4 & 0 & 0 \\
\bottomrule
\end{tabular}}
\caption{Missing answers by kind for every cell with at least three (all counted as failures in every accuracy): \emph{refusal}, a text reply declining the task; \emph{text only}, an answer written in the message instead of the tool call; \emph{other tool} and \emph{multiple calls}, calls to a non-existent tool or more than one call; \emph{stream error}, a provider-side error after the request was accepted; \emph{truncated}, the 16,384-token output limit reached (for Anthropic thinking runs, entirely in thinking); \emph{empty}, no content.}
\label{tab:missing}
\end{table}

\begin{table}[htbp]
\centering\scriptsize
\setlength{\tabcolsep}{3pt}
\begin{tabular}{lllrr}
\toprule
Cohort & Model, effort & Arm & Automatic $-$ forced (future joint) & Automatic $-$ forced (stock exact) \\
\midrule
Synthetic & GPT-5.4 mini, low & raw & +19.3 [+11.5, +27.1] & +8.3 [+1.6, +15.1] \\
Synthetic & GPT-5.4 mini, high & raw & +54.2 [+44.3, +63.5] & +40.6 [+31.8, +49.5] \\
Synthetic & GPT-5.4 mini, low & canonical & +10.4 [+3.1, +17.7] & +3.6 [-1.6, +8.9] \\
Synthetic & GPT-5.4 mini, high & canonical & +58.3 [+49.5, +66.7] & +40.6 [+31.8, +49.5] \\
Synthetic & GPT-5.4, low & raw & +10.4 [+5.2, +16.1] & +7.3 [+2.6, +12.5] \\
Synthetic & GPT-5.4, high & raw & +4.7 [+1.0, +8.3] & +2.6 [+0.0, +5.2] \\
Synthetic & GPT-5.4, low & canonical & +0.0 [-2.6, +2.6] & -1.0 [-3.1, +1.0] \\
Synthetic & GPT-5.4, high & canonical & +5.2 [+2.1, +8.9] & +1.6 [+0.0, +3.6] \\
Synthetic & gpt-oss-20b, low & raw & +0.5 [-5.7, +6.8] & +0.5 [-6.2, +7.3] \\
Synthetic & gpt-oss-20b, high & raw & -3.1 [-8.9, +2.1] & -1.6 [-6.8, +3.6] \\
Synthetic & gpt-oss-20b, low & canonical & -1.0 [-8.3, +6.2] & +4.2 [-2.1, +10.4] \\
Synthetic & gpt-oss-20b, high & canonical & -6.8 [-13.5, +0.0] & -5.7 [-12.5, +1.0] \\
Synthetic & gpt-oss-120b, low & raw & +3.6 [-3.1, +10.4] & +1.0 [-4.7, +6.8] \\
Synthetic & gpt-oss-120b, high & raw & -4.7 [-8.9, -1.0] & -4.7 [-8.9, -1.0] \\
Synthetic & gpt-oss-120b, low & canonical & +4.2 [-3.1, +12.0] & -1.6 [-7.3, +4.2] \\
Synthetic & gpt-oss-120b, high & canonical & +2.6 [-2.6, +7.8] & +2.6 [-2.6, +7.8] \\
Synthetic & Qwen3.8-27B, low & raw & +18.8 [+12.5, +25.5] & +18.2 [+12.0, +25.0] \\
Synthetic & Qwen3.8-27B, high & raw & +8.3 [+1.6, +15.1] & +6.8 [+0.0, +13.5] \\
Synthetic & Qwen3.8-27B, low & canonical & +7.8 [+4.2, +12.0] & +7.8 [+4.2, +12.0] \\
Synthetic & Qwen3.8-27B, high & canonical & +1.0 [-4.7, +6.8] & +1.6 [-3.1, +5.7] \\
Synthetic & DeepSeek V4 Flash, low & raw & +96.4 [+93.8, +99.0] & +92.2 [+88.0, +95.8] \\
Synthetic & DeepSeek V4 Flash, high & raw & +91.7 [+87.0, +95.8] & +87.5 [+81.8, +92.7] \\
Synthetic & DeepSeek V4 Flash, low & canonical & +96.9 [+94.3, +99.0] & +91.1 [+87.0, +94.8] \\
Synthetic & DeepSeek V4 Flash, high & canonical & +96.4 [+93.2, +99.0] & +94.3 [+90.1, +97.9] \\
Synthetic & GLM-5.3 Flash, low & raw & +3.1 [-6.2, +12.5] & -2.1 [-9.4, +4.7] \\
Synthetic & GLM-5.3 Flash, high & raw & +2.1 [-5.2, +9.4] & +1.0 [-3.6, +5.7] \\
Synthetic & GLM-5.3 Flash, low & canonical & +1.6 [-6.8, +9.9] & +1.6 [-5.2, +8.3] \\
Synthetic & GLM-5.3 Flash, high & canonical & +2.1 [-4.7, +8.9] & +2.1 [-2.1, +6.2] \\
Synthetic & DeepSeek V4 Pro, low & raw & +2.6 [-0.5, +5.7] & +3.1 [+0.0, +6.2] \\
Synthetic & DeepSeek V4 Pro, high & raw & +1.6 [-2.6, +5.7] & +2.1 [-1.6, +5.7] \\
Synthetic & DeepSeek V4 Pro, low & canonical & +0.5 [-1.6, +2.6] & +1.0 [+0.0, +2.6] \\
Synthetic & DeepSeek V4 Pro, high & canonical & +15.1 [+9.4, +20.8] & +14.1 [+8.3, +19.8] \\
Retail profiles & GPT-5.4 mini, low & raw & +2.1 [-6.2, +11.5] & +4.2 [-4.2, +13.5] \\
Retail profiles & GPT-5.4 mini, high & raw & +28.1 [+15.6, +39.6] & +18.8 [+7.3, +30.2] \\
Retail profiles & GPT-5.4 mini, low & canonical & +35.4 [+22.9, +47.9] & +2.1 [-6.2, +10.4] \\
Retail profiles & GPT-5.4 mini, high & canonical & +65.6 [+55.2, +75.0] & +61.5 [+50.0, +71.9] \\
Retail profiles & GPT-5.4, low & raw & +7.3 [+3.1, +12.5] & +7.3 [+3.1, +12.5] \\
Retail profiles & GPT-5.4, high & raw & +1.0 [+0.0, +3.1] & +1.0 [+0.0, +3.1] \\
Retail profiles & GPT-5.4, low & canonical & +8.3 [+3.1, +13.5] & +0.0 [+0.0, +0.0] \\
Retail profiles & GPT-5.4, high & canonical & +11.5 [+6.2, +17.7] & +9.4 [+4.2, +15.6] \\
Retail profiles & gpt-oss-20b, low & raw & -1.0 [-3.1, +0.0] & -1.0 [-7.3, +4.2] \\
Retail profiles & gpt-oss-20b, high & raw & +0.0 [+0.0, +0.0] & +0.0 [+0.0, +0.0] \\
Retail profiles & gpt-oss-20b, low & canonical & +6.2 [-3.1, +15.6] & +2.1 [-8.3, +12.5] \\
Retail profiles & gpt-oss-20b, high & canonical & -1.0 [-14.6, +12.5] & +0.0 [-13.5, +13.5] \\
Retail profiles & gpt-oss-120b, low & raw & -2.1 [-8.3, +3.1] & -7.3 [-13.6, -1.0] \\
Retail profiles & gpt-oss-120b, high & raw & +0.0 [+0.0, +0.0] & +0.0 [+0.0, +0.0] \\
Retail profiles & gpt-oss-120b, low & canonical & -3.1 [-14.6, +9.4] & -2.1 [-11.5, +8.3] \\
Retail profiles & gpt-oss-120b, high & canonical & -1.0 [-10.4, +8.3] & -1.0 [-10.4, +8.3] \\
Retail profiles & DeepSeek V4 Flash, low & raw & +94.8 [+88.5, +99.0] & +87.5 [+79.2, +94.8] \\
Retail profiles & DeepSeek V4 Flash, high & raw & +93.8 [+87.5, +99.0] & +89.6 [+80.2, +96.9] \\
Retail profiles & DeepSeek V4 Flash, low & canonical & +95.8 [+91.7, +99.0] & +87.5 [+79.2, +94.8] \\
Retail profiles & DeepSeek V4 Flash, high & canonical & +97.9 [+94.8, +100.0] & +88.5 [+80.2, +95.8] \\
Retail profiles & GLM-5.3 Flash, low & raw & -5.2 [-18.8, +8.3] & -2.1 [-15.6, +11.5] \\
Retail profiles & GLM-5.3 Flash, high & raw & +0.0 [-11.5, +11.5] & -3.1 [-11.5, +5.2] \\
Retail profiles & GLM-5.3 Flash, low & canonical & -6.2 [-20.8, +7.3] & -3.1 [-12.5, +5.2] \\
Retail profiles & GLM-5.3 Flash, high & canonical & +1.0 [-8.3, +10.4] & +0.0 [-6.2, +6.2] \\
Retail profiles & DeepSeek V4 Pro, low & raw & +4.2 [-3.1, +11.5] & +6.2 [+0.0, +12.5] \\
Retail profiles & DeepSeek V4 Pro, high & raw & +17.7 [+9.4, +26.0] & +17.7 [+9.4, +26.0] \\
Retail profiles & DeepSeek V4 Pro, low & canonical & +3.1 [+0.0, +7.3] & +2.1 [+0.0, +5.2] \\
Retail profiles & DeepSeek V4 Pro, high & canonical & +21.9 [+14.6, +29.2] & +21.9 [+14.6, +29.2] \\
\bottomrule
\end{tabular}
\caption{Interface effect: automatic tool choice minus forced answer call at fixed model, effort, input and history; paired differences in percentage points with descriptive 95\% history-bootstrap intervals.}
\label{tab:interface}
\end{table}

%% file: interaction_raw_table.tex
\begin{table}[t]
\centering\footnotesize
\setlength{\tabcolsep}{3.5pt}
\begin{tabular}{lrrrrrrr}
\toprule
 & \multicolumn{3}{c}{Forced answer call} & \multicolumn{3}{c}{Automatic tool choice} & \\
\cmidrule(lr){2-4}\cmidrule(lr){5-7}
Model & low & high & high$-$low & low & high & high$-$low & Interaction \\
\midrule
GPT-5.4 mini & 37.5 & 25.0 & -12.5 & 56.8 & 79.2 & +22.4 & +34.9 [+22.4, +46.9] \\
GPT-5.4 & 88.0 & 94.3 & +6.2 & 98.4 & 99.0 & +0.5 & -5.7 [-12.5, +1.0] \\
gpt-oss-20b & 23.4 & 42.2 & +18.8 & 24.0 & 39.1 & +15.1 & -3.6 [-12.0, +4.7] \\
gpt-oss-120b & 45.8 & 53.6 & +7.8 & 49.5 & 49.0 & -0.5 & -8.3 [-16.7, -0.5] \\
Qwen3.8-27B & 76.6 & 80.2 & +3.6 & 95.3 & 88.5 & -6.8 & -10.4 [-20.3, -0.5] \\
DeepSeek V4 Flash & 2.1 & 1.6 & -0.5 & 98.4 & 93.2 & -5.2 & -4.7 [-9.9, +0.5] \\
GLM-5.3 Flash & 52.6 & 74.5 & +21.9 & 55.7 & 76.6 & +20.8 & -1.0 [-13.0, +10.9] \\
DeepSeek V4 Pro & 96.4 & 91.7 & -4.7 & 99.0 & 93.2 & -5.7 & -1.0 [-6.2, +3.6] \\
\bottomrule
\end{tabular}
\caption{Requested reasoning effort $\times$ tool-choice interface, synthetic cohort, raw input, future-joint accuracy (\%; 192 calls per cell). high$-$low: paired effort effect within an interface. Interaction: automatic-choice effort effect minus forced-call effort effect, paired by history, with a descriptive 95\% history-bootstrap interval (10,000 draws); a positive interaction means the forced call turns extra effort into a liability. Requested effort is a family-specific provider setting, not a compute match (observed reasoning tokens: Appendix~\ref{app:ladder}).}
\label{tab:interaction_raw}
\end{table}

%% file: extension_state_table.tex
\begin{table}[t]
\centering\footnotesize
\setlength{\tabcolsep}{4pt}
\begin{tabular}{lrrrrrrr}
\toprule
 & \multicolumn{3}{c}{Raw input} & \multicolumn{3}{c}{Canonical input} & Snapshot \\
Cohort, configuration & model & hybrid & inside & model & hybrid & inside & model \\
\midrule
Synthetic, GPT-5.4 mini (low) & 37.5 & 55.2 & 91.7 & 76.0 & 93.8 & 97.9 & 89.6 \\
Synthetic, GPT-5.4 (low) & 88.0 & 91.7 & 100.0 & 98.4 & 98.4 & 97.4 & 99.0 \\
Synthetic, Sonnet 4.6 & 99.0 & 100.0 & -- & 99.5 & 100.0 & -- & 100.0 \\
Synthetic, Sonnet + thinking & 97.9 & 100.0 & -- & 100.0 & 100.0 & -- & 100.0 \\
\midrule
Retail profiles, GPT-5.4 mini (low) & 12.5 & 17.7 & 89.6 & 46.9 & 99.0 & 100.0 & 89.6 \\
Retail profiles, GPT-5.4 (low) & 90.6 & 90.6 & 100.0 & 91.7 & 99.0 & 100.0 & 99.0 \\
Retail profiles, Sonnet 4.6 & 95.8 & 95.8 & -- & 99.0 & 100.0 & -- & 100.0 \\
Retail profiles, Sonnet + thinking & 97.9 & 100.0 & -- & 100.0 & 100.0 & -- & 100.0 \\
\bottomrule
\end{tabular}
\caption{Future-joint accuracy (\%, all three scenario outputs exact) by input and code placement; 192 synthetic and 96 profile calls per cell; invalid answers fail. \emph{model}: all five fields from the model; \emph{hybrid}: model returns the two historical fields, code applies the schedule (equals reconstruction accuracy by conservation); \emph{inside}: code-execution tool available during the call (OpenAI only); \emph{snapshot}: exact state supplied, model rolls out; dashes mark arms not run. All-five accuracy: Appendix~\ref{app:tables}.}
\label{tab:state}
\end{table}

%% file: extension_effort_table.tex
\begin{table}[t]
\centering\footnotesize
\begin{tabular}{llrrrrrr}
\toprule
 & & \multicolumn{3}{c}{Synthetic} & \multicolumn{3}{c}{Retail profiles} \\
Model & Arm & low & medium & high & low & medium & high \\
\midrule
GPT-5.4 mini, forced call & Raw & 37.5 & 26.6 & 25.0 & 12.5 & 16.7 & 20.8 \\
 & Canonical & 76.0 & 46.9 & 34.9 & 46.9 & 38.5 & 29.2 \\
 & Snapshot & 89.6 & 97.9 & 97.4 & 89.6 & 94.8 & 93.8 \\
\midrule
GPT-5.4, forced call & Raw & 88.0 & 95.3 & 94.3 & 90.6 & 95.8 & 99.0 \\
 & Canonical & 98.4 & 95.8 & 94.8 & 91.7 & 94.8 & 88.5 \\
\midrule
GPT-5.4 mini, auto choice & Raw & 56.8 & -- & 79.2 & 14.6 & -- & 49.0 \\
 & Canonical & 86.5 & -- & 93.2 & 82.3 & -- & 94.8 \\
\bottomrule
\end{tabular}
\caption{Future-joint accuracy (\%) by requested reasoning effort, identical prompts and tasks; 192 synthetic and 96 profile calls per cell; the forced-call low columns are the original cohorts. Forced call: the answer function is the only permitted action; auto choice: the model may respond freely and a missing answer fails. Dashes: not run.}
\label{tab:effort}
\end{table}